\documentstyle[12pt,twoside,amsfonts,oldlfont]{article}

\normalbaselines

\makeatletter
\@addtoreset{equation}{section}
\makeatother

\newcommand{\beq}{\begin{equation}}
\newcommand{\eeq}{\end{equation}}
\font\goth=eufm10 scaled \magstep1

\def\pint{\mathbin{\raise1.5pt\hbox{$\underline{\raise-.5pt\hbox
{$\phantom{n}$}}\mskip-2mu\scriptstyle|$}}}
\def\vtaub{V(\tau,b)}
\def\tvtaub{\widetilde{V}(\tau, b)}
\def\dtau{D(\tau)}
\def\demi{{1\over2}}
\def\osp{\hbox{osp}(2/2)}
\def\ospr{\hbox{osp}(2/2,\R)}
\def\ospc{\hbox{osp}(2/2,\C)}
\def\OSp{\hbox{OSp}(2/2)}
\def\ssu{\hbox{su}(1,1)}
\def\su{\hbox{SU}(1,1)}
\def\du{{\cal D}^{(1)}}
\def\lu{{\cal L}^{(1)}}
\def\dud{{\cal D}^{(1\vert 2)}}
\def\duu{{\cal D}^{(1\vert 1)}}
\def\lud{{\cal L}^{(1\vert 2)}}
\def\ludp{\lud_{\rm p}}
\def\om{{\cal O}_M}
\def\am{{\cal A}_M}
\def\de{1-|z|^2}
\def\vb{\bar z}
\def\ko{|0\rangle}

\def\pv{{\partial\over\partial z}}
\def\pvb{{\partial\over\partial \vb}}
\def\t{\theta}
\def\tb{\bar \theta}
\def\chib{\bar \chi}
\def\ket{|z,\theta,\chi\rangle}
\def\ketb{|\vb,\tb,\chib\rangle}

\def\brab{\langle \zb,\tb,\chib}
\def\pt{{\partial\over\partial\theta}}
\def\ptb{{\partial\over\partial\vphantom{\bar{\bar\theta}}\tb}}
\def\pchi{{\partial\over\partial \chi}}
\def\pchib{{\partial\over\partial \chib}}
\def\lw{{ |b,\tau,\tau\rangle }}
\def\h{{\cal H}}
\def\I{{\Bbb I}}
\def\L{{\Bbb L}}
\def\C{{\Bbb C}}
\def\R{{\Bbb R}}
\def\Z{{\Bbb Z}}
\def\N{{\Bbb N}}
\def\E{{\Bbb E}}
\def\F{{\Bbb F}}
\def\n{\hbox{\goth n}}
\def\b{\hbox{\goth b}}
\def\hh{\hbox{\goth h}}
\def\g{\hbox{\goth g}}
\def\mz{|z|^2}
\def\zb{\bar z}
\def\pz{{\partial\over\partial z}}
\def\pzb{{\partial\over\partial \vb}}
\def\nil{{\cal N}}
\def\she{{\cal E}}
\def\aud{{\cal A}^{(1|2)}}
\def\tu{\theta^1}
\def\td{\theta^2}
\def\tub{\bar\theta^1}
\def\tdb{\bar\theta^2}
\font\sect=cmbx10 scaled\magstep2

\begin{document}
%%%%%%%%%%%%%%%%%%%%%%%%%%%%%%%%%%
\renewenvironment{thebibliography}[1]
  {
\section*{References}\addcontentsline{toc}{section}{References}
    \begin{list}{\arabic{enumi}.}
    {\usecounter{enumi} \setlength{\parsep}{0pt}
     \setlength{\itemsep}{3pt} \settowidth{\labelwidth}{#1.}
     \sloppy
    }}{\end{list}}
%%%%%%%%%%%%%%%%%%%%%%%%%%%%%%%%%%%
\renewcommand{\thefootnote}{\fnsymbol{footnote}}
\parindent=1.5pc
%%%%%%%%%%%%%%%%%%%%%%%%%%%%%%%%%%%
%%%%%%%%%%%%%%%%%%%%%%%%%%%%%%%%%%%
{\noindent March 15, 1994  \hfill CRM-1876}

\vglue 1.5cm
\begin{center}
{{\sect
SUPERCOHERENT STATES, SUPER K\"AHLER\\
\vglue 3pt
\vglue 3pt
GEOMETRY AND GEOMETRIC QUANTIZATION}\\

\vglue 1.0cm
{\bf Amine M. El Gradechi\,$^{1,
2,}$\footnote[1]{\rm E-mail:
elgradec@ere.umontreal.ca},  Luis M.
Nieto\,$^{1,}$\footnote[2]{\rm E-mail:
nietol@ere.umontreal.ca}}\\
\vglue 0.3cm
\baselineskip=14pt {{$^{1)}\,$}\it Centre de Recherches
Math\'ematiques, Universit\'e de Montr\'eal}\\
\baselineskip=14pt {\it C.P. 6128-centre-ville, Montr\'eal (Qu\'ebec) H3C
3J7, Canada}\\
\vglue 0.5cm {{$^{2)}\,$}\it Department of Mathematics and Statistics,
Concordia University}\\
\baselineskip=14pt {\it Montr\'eal (Qu\'ebec) H4B 1R6, Canada} \\
\vglue 1.cm {ABSTRACT}}
\end{center}
\vglue 0.2cm
{\footnotesize{ \begin{quote}
  \noindent
Generalized coherent states provide a means of connecting square
integrable representations of a semi-simple Lie group with the symplectic
geometry of some of its homogeneous spaces.  In the first part of the
present work this point of view is extended to the  supersymmetric context,
through the study of the \OSp\  coherent states.  These are explicitly
constructed starting from the known  abstract typical and atypical
representations of  \osp.  Their underlying geometries turn out to be
those of supersymplectic \OSp-homogeneous spaces.  Moment maps
identifying the latter with coadjoint orbits of \OSp\ are exhibited via
Berezin's symbols.  When considered within Rothstein's general paradigm,
these results lead to a natural general definition of a super K\"ahler
supermanifold, the supergeometry of which is determined in terms of the
usual geometry of holomorphic Hermitian vector bundles over K\"ahler
manifolds.  In particular, the supergeometry of the above orbits is
interpreted in terms of the geometry of Einstein-Hermitian vector bundles.
In the second  part, an extension of the full geometric quantization
procedure is applied to the same coadjoint orbits.  Thanks to the super
K\"ahler character of the latter, this procedure leads to explicit super
unitary irreducible representations of \OSp\ in super Hilbert spaces of
$L^2$ superholomorphic sections of prequantum bundles of the Kostant
type.  This work lays the foundations of a program aimed at  classifying
Lie supergroups' coadjoint orbits and their associated irreducible
representations, ultimately leading to harmonic superanalysis. For
this purpose a set of consistent conventions is exhibited.
\end{quote} }}

\thispagestyle{empty}
\vglue .6cm
{\footnotesize
\begin{description}
\item[Key-Words:] Coherent states, supersymplectic
geometry, super K\"ahler supermanifolds, geometric quantization,
Lie superalgebras
\item[1991 MSC:] 17A70, 32C11, 53C55, 58A50, 58F06, 81R30, 81S10
\item[PACS:] 02.20 Sv, 02.40, 03.65 Sq
\end{description}}

\vfill\eject
\vglue 3cm
\centerline{R\'ESUM\'E}

\vglue 0.3cm

{\footnotesize{ \begin{quote}
\noindent
Les \'etats coh\'erents g\'en\'eralis\'es fournissent un moyen de relier les
repr\'esentations de carr\'e int\'egrables d'un groupe de Lie semi-simple
\`a la g\'eom\'etrie symplectique de certains de ses espaces homog\`enes.
Dans la premi\`ere partie de ce travail, l'extension supersym\'etrique de ce
point de vue est consid\'er\'ee \`a travers l'\'etude des \'etats coh\'erents
de $\OSp$.  Ceux-ci sont explicitement construits \`a partir des
repr\'esentations abstraites typique et atypique connues de $\osp$.  Les
g\'eom\'etries associ\'ees \`a ces \'etats se r\'ev\'elent \^etre celles
d'espaces homog\`enes supersymplectiques de $\OSp$.  Des applications
moments identifiant ces derniers avec des orbites coadjointes de $\OSp$
sont produites via les symboles de Berezin.  Interpr\'et\'es dans le cadre
de la th\'eorie g\'en\'erale d\'evelopp\'ee par Rothstein, ces r\'esultats
conduisent \`a une d\'efinition g\'en\'erale et naturelle d'une
supervari\'et\'e super K\"ahlerienne.  La superg\'eom\'etrie de celle-ci
est d\'etermin\'ee en termes de la g\'eom\'etrie habituelle de fibr\'es
vectoriels holomorphes Hermitiens au dessus de vari\'et\'es
K\"ahleriennes.  En particulier, la superg\'eom\'etrie des orbites
mentionn\'ees ci-dessus  s'interpr\'etent en termes de fibr\'es
d'Einstein-Hermitiens. Dans la seconde partie, une extension de la
proc\'edure compl\`ete de quantification g\'eom\'etrique est appliqu\'ee
aux m\^emes orbites coadjointes.  Leur caract\`ere super K\"ahlerien
permet alors de produire des repr\'esentations super unitaires
irr\'eductibles explicites de $\OSp$ dans des super espaces de Hilbert de
sections $L^2$ superholomorphes de fibr\'es pr\'equantiques du type
d\'ecrit par Kostant.  Ce travail pose les premiers jalons  d'un
programme ayant pour objectif de classifier les orbites coadjointes des
supergroupes de Lie et les repr\'esentations qui leurs sont associ\'ees,
pour finalement aboutir \`a une superanalyse harmonique.  Pour cela un
ensemble consistent de conventions est pr\'esent\'e.
\end{quote} }}
\thispagestyle{empty}
\vfill\eject

%%%%%%%%%%%%%%%%%%%%%%%%%%%%%%%%%%%%
%%%%%%%%%% 0. Contents %%%%%%%%%%%%%%%%%%%
%%%%%%%%%%%%%%%%%%%%%%%%%%%%%%%%%%%%
{\baselineskip=12pt\footnotesize
\tableofcontents}
\thispagestyle{empty}

\vfill\eject

%%%%%%%%%%%%%%%%%%%%%%%%%%%%%%%%%%%%
%%%%%%%%%% 1. Introduction %%%%%%%%%%%%%%%%%
%%%%%%%%%%%%%%%%%%%%%%%%%%%%%%%%%%%%
\baselineskip=18pt
\setcounter{page}{1}
%%%%%%%%%%%%%%%%%%%%%%%%%%%%%%%%%%%%
\section{Introduction}\label{sec1}

\noindent{\bf1.1.}  Coherent states were originally those very special
quantum states of the harmonic oscillator first introduced by Schr\"odinger
\cite{Schro} and studied further by Glauber \cite{Glauber}.  Their
special character stems from their property of being the closest possible
states to the classical theory.  This is reflected  in the fact that they
minimize  Heisenberg's uncertainty relations.  Moreover, they form an
overcomplete basis of the Hilbert space of quantum states.  Since
Glauber's contribution, the concept of coherent states has evolved very
rapidly.  The key step of this evolution was without any doubt Perelemov's
group theoretical generalization \cite{Pere}.
Nowadays' coherent states find applications in several areas of
physics and mathematics.  Let us mention, for instance, their
occurrence in quantum optics \cite{KlaSka},  in signal analysis (where they
are called wavelets)
\cite{Daub}, and in mathematical physics (in connection with the
quantization-versus-classical-limit procedures) \cite{Berezin1,Onofri}
(see also \cite{Pere} and references therein).  This last
application constitutes the main interest of the present paper.  More
precisely,  we extend to the supersymmetric context both the
coherent states approach to the evaluation of the classical limit, and the
geometric quantization of coadjoint orbits.  Even though these two points
have already been studied in the case of the OSp$(1/2)$ Lie supergroup in
\cite{Amine1,Amine2}, the analysis of a less trivial example, such as the
\OSp\ Lie supergroup considered here, brings new insights that improve
our understanding of the very interesting supergeometric structures
underlying the so-called supercoherent states.   Let us now present an
overview of the subject.  More details and references about the
coherent states methods can be found in \cite{Pere,KlaSka,Gilmore}.

\smallskip
\noindent{\bf1.2.}  The harmonic oscillator coherent states admit a group
theoretical construction based on the Weyl-Heisenberg group underlying
this physical system.  By extending this construction to general Lie groups,
Perelomov introduced the notion of generalized coherent states
\cite{Pere}.  For a given Lie group $G$ and a given unitary irreducible
representation $U$ of $G$ in some Hilbert space $\h$, the generalized
coherent states are the states belonging to a
$U(G)$-orbit in $\h$ through a chosen initial state, usually
called the fiducial state.  The generalized coherent states so defined are
too general to share the interesting properties of the harmonic oscillator
ones.  Indeed, the overcompleteness property holds only when $U$ is a
square integrable representation, and the uncertainty relations
associated with the Lie algebra $\g$ of $G$ are minimized only when
the fiducial state is a highest weight state
\cite{Pere,Delb,dbe}.  In what follows we will only consider this
type of generalized coherent states and we will simply designate them by
coherent states (CS) or $G$-coherent states ($G$-CS).

\smallskip
\noindent{\bf1.3.}  Not all Lie groups possess square integrable
representations.  Unfortunately, this is the case for some physically
relevant groups, such as the Poincar\'e group.  There is however enough
room to actually take advantage of the particularly rich properties of
the generalized CS.  Indeed, all representations of compact semi-simple
Lie groups, and all discrete series representations of the non-compact
semi-simple Lie groups, are square integrable. Moreover, physical
interpretations can be attached to the associated CS.  For instance,
SU$(2)$-CS \cite{Ratc} allow a semi-classical description of spin and the
SU$(1,1)$-CS
\cite{dbe} are optimally localized states for the quantum mechanics of a
free particle on the $(1+1)$-dimensional anti-de~Sitter spacetime.  It is
worth mentioning that the square integrability condition  has recently
been replaced by the less restrictive notion of square integrability
modulo a subgroup, which allowed the construction of (quasi-)coherent
states for the Poincar\'e group in $(1+1)$-dimensions
\cite{AAG1}.  A more general framework is
described in \cite{AAG2}.

\smallskip
\noindent{\bf1.4.}
Let us now discuss the relevance of the CS to the
quantization-versus-classical-limit procedures.  A classical
$G$-elementary system is generally described  by a coadjoint orbit of
$G$, which is a symplectic
$G$-homogeneous space $(G/H,\omega)$, where $H$  is a closed
subgroup of $G$, and $\omega$ is a $G$-invariant,
closed and non-degenerate $2$-form on $G/H$.  On the other hand, a
quantum $G$-elementary system is described
by a pair $(U,\h)$, where $U$ is a
Unitary Irreducible Representation (UIR) of $G$ in a Hilbert space $\h$.
Classical and quantum $G$-elementary systems are related to each
other by on the one hand, the quantization methods, such as the
Kirillov-Kostant-Souriau geometric quantization (also known as
the orbits method) \cite{Kiri1,Kost1,Souriau}, or Berezin's
quantization \cite{Berezin1}, or the deformation quantization
\cite{Lichne}, and on the other hand the classical limit procedures, such
as the CS-inspired one described by Onofri \cite{Onofri}.  Since at
different stages of this work we will be dealing with these two kinds of
procedures, we now briefly hint at their intuitive content.

\smallskip
\noindent{\bf1.5.}  Whenever
$U$ is a square integrable UIR of $G$, one can construct a family of CS
parametrized  by $G/H$, where $H$ is the closed subgroup of $G$ which
leaves invariant, up to a phase, the fiducial state.  By construction, this
family bears in its very structure enough information to allow
one to equip
$G/H$ with a
$G$-invariant symplectic form $\omega$, which makes $(G/H,\omega)$
into the classical $G$-elementary system
describing the classical limit of the quantum $G$-elementary system
$(U,\h)$.  More precisely, this explicit construction leads to a K\"ahler
$G$-homogeneous space \cite{Onofri}.  Combining this derivation with the
evaluation of Berezin's covariant symbols \cite{Berezin1} provides one
with a moment map that identifies $(G/H,\omega)$ with a coadjoint orbit
of $G$.  Berezin's symbols are the mean values of the quantum
generators of $\g$ in the CS.

\smallskip
\noindent{\bf1.6.} Obviously, evaluating the classical limit is
more natural than quantizing.  However, because of their better
understanding of classical theories, mathematical-physicists are very
much interested in having a quantization procedure that would allow
the translation of some of the well established classical physical
understanding to a quantum counterpart.  Several quantization procedures
are now available. Three of them have already been mentioned.
Geometric quantization is the method that we will be dealing with in this
paper.  In its simplest version, this technique associates a quantum
$G$-elementary system to a  classical one
$(G/H,\omega)$, provided that $[\omega]$ is an integral
cohomology  class
and that $G/H$ admits an invariant polarization.

\smallskip
\noindent{\bf1.7.}  The very natural question we address in
this work can be formulated as follows: {\it How do the
quantization-versus-classical-limit procedures depicted above extend to
the supersymmetric context?\/}  We provide an answer to this, by
studying very specific though non-trivial examples, namely those of the
typical and the atypical
\OSp-elementary systems.  To shed more light on our motivations, we
now situate our present contribution within the framework of the fast
developing field of supermathematics.

\smallskip
\noindent{\bf1.8.}  Supermathematics is the collection of mathematical
tools developed during the last thirty years in order to provide
physicists with a rigorous framework for the study of the so-called
supersymmetric theories, such as supergravity and superstrings.
These are
theories that possess symmetries which mix their bosonic and fermionic
degrees of freedom.  Some of the tools were already available before
these theories really triggered the interest of a large number of
researchers (see \cite{Freund} pp.~26--28).  A description of the super
extensions
of the usual analytic, algebraic and geometric concepts that have so far
been obtained would unfortunately lead us too far from the subject of
this paper; we refer the interested reader to the existing literature
\cite{Berezin2}--\cite{Gijs0} and  confine  our description to those super
ingredients that are crucial for answering the question raised in 1.7.

\smallskip
\noindent{\bf1.9.}  Contrary to
Lie superalgebras, there exist several notions of Lie supergroups which
correspond to the different definitions of a supermanifold (see \cite{Bartocci}
for more details).  Concerning the representation theory,
only abstract representations (Verma modules) of some simple Lie
superalgebras are known
\cite{Berezin3,Scheunert,Kac,ShNaRitt,Nish}.  Using these representations,
supercoherent states have recently been explicitly constructed
\cite{Balal1,Balal2,Fatal}.  On the other hand, supergeometric
concepts such as supersymplectic supermanifolds and super coadjoint
orbits have been understood since the end of the 70s
\cite{Berezin3,Kost2} (see \cite{Duval} for a recent application in
physics).  A very nice characterization of the latter in terms of
usual  geometric objects has been obtained by Rothstein
\cite{Roth1} (see also \cite{Monterde}).  Hence, all the  ingredients
needed for answering the question formulated in 1.7 are available.  It
remains to super extend the methods described in 1.5 and 1.6.  This is
explicitly  carried out here for the case of the typical and
atypical \OSp-CS.

\smallskip
\noindent{\bf1.10.}  The paper is organized as follows: In Section~2 the
so-called typical and atypical super unitary irreducible representations of
the Lie superalgebra $\osp$,  which super extend the discrete series
representations of its subalgebra
\ssu, are described. Then, the typical coherent states are constructed.  By
doing this we reproduce the construction of the \OSp-CS obtained in
\cite{Balal2}, and cure some of the discrepancies appearing there.  In
Section~3 a super extension of the methods depicted in 1.5 is then applied
to these CS.  Hence, the \OSp-homogeneous superspace parametrizing the
latter, which we denote by $\dud$,  is explicitly equipped with an
\OSp-invariant supersymplectic structure~$\omega$.  Moreover, the
evaluation of Berezin's covariant symbols exhibits a momentum map that
identifies
$(\dud,\omega)$ with an
\OSp-coadjoint orbit.  Finally, the invariant super measure on $\dud$
obtained in
\cite{Balal2} is recovered through a simpler computation.  In Section~4,
after an introduction to the general theory,
the supergeometry of
$(\dud,\omega)$ is studied further.  As a result,
$(\dud,\omega)$ is shown to be not only a non-trivial example of
Rothstein's general supersymplectic supermanifolds \cite{Roth1} but also
a non-trivial example of the notion of a super K\"ahler supermanifold
already discussed in \cite{Amine1,Amine2}.  In this particular setting, we
show how Rothstein's characterization can be improved.  In
Section~5, after a brief introduction to the usual geometric quantization
procedure, a super extension of it is applied to
$(\dud,\omega)$.  General superprequantization has been
developed by Kostant
\cite{Kost2}, however the lack of a notion of polarization prevented him from
completing the quantization.  Here, the super K\"ahler character of
$(\dud,\omega)$ singles out a natural invariant super K\"ahler
polarization which leads to the complete quantization.  The  coherent
states associated to the atypical representations (atypical CS), their
underlying supergeometry, and its quantization are described in Section~6.
In particular the atypical coadjoint orbit is identified.  Section~7 gathers
additional results and discussions, while concluding remarks and
possible extensions of the present work are displayed in Section~8.
Our conventions and notations are presented in Appendix~A, while useful
constructions are relegated to Appendix~B.

\smallskip
\noindent{\bf1.11.} Before presenting the details, we would like
to make clear some important points concerning our approach and strategy.
Perelomov's construction of $G$-CS can be explicitly carried
out starting simply from an infinitesimal version of a UIR of $G$
\cite{Pere}.  This approach,  used in \cite{Balal1,Balal2}, is also
adopted here.  It might also seem strange that we start with a
representation of the Lie superalgebra, evaluate its classical limit
through the associated CS and then quantize the obtained supersymplectic
supermanifold in order to construct a representation of the same Lie
superalgebra!  In fact, the representation we start with is, as we will see,
an abstract representation (Verma module), while the second is  an
explicit one.  The latter, is realized in a super Hilbert space of
superholomorphic sections of a line bundle sheaf over the considered
coadjoint orbit.  This representation is an important step towards
constructing explicit super UIR of Lie supergroups.

%%%%%%%%%%%%%%%%%%%%%%%%%%%%%%%%%%%%
%%%%%%%%%% 2. IR and SCS %%%%%%%%%%%%%%%%%%
%%%%%%%%%%%%%%%%%%%%%%%%%%%%%%%%%%%%

\section{osp(2/2) representations and typical OSp(2/2)-CS}\label{sec2}

\noindent This section is devoted  to two main purposes.  We first
describe the $\osp$ Lie superalgebra and its lowest
weight typical and atypical representations, then we construct the
associated coherent states.

\subsection{osp(2/2) representations}

The superalgebra we consider here is the real orthosymplectic
superalgebra $\ospr$.  Throughout, we will simply denote it $\osp$.  It is
a real non-compact form of the basic classical simple Lie superalgebra
$\ospc$, which corresponds to $C(2)$ in Kac's classification \cite{Kac}.  Its
$\Z_2$-grading, $\osp=\osp_{\bar 0}\oplus\osp_{\bar 1}$, is such that the
even component $\osp_{\bar 0}=\hbox{so}(2)\oplus \hbox{sp}(2,\R)$, and
the odd component
$\osp_{\bar 1}$ is an $\osp_{\bar 0}$-module.  In what follows, using the
isomorphism $\hbox{sp}(2,\R)\cong\ssu$, we will consider $\ssu$ instead
of $\hbox{sp}(2,\R)$.

As for Lie algebras, the construction of representations of $\osp$
relies on its complexification $\ospc$.  We now display a description of
the latter superalgebra.  Since
$\ospc$ is of type~I \cite{Kac,Cornwell}, its odd component $\ospc_{\bar
1}$ decomposes into two irreducible
$\ospc_{\bar 0}$-modules, namely,
$\ospc_{\bar 1}=\ospc_{-1}\oplus \ospc_{1}$.
Moreover $\ospc$ admits the following $\Z_2$-compatible $\Z$-gradation:
\beq
\ospc=\ospc_{-1}\oplus \ospc_0\oplus
\ospc_1, \label{2.1}
\eeq
where $\ospc_0=\ospc_{\bar 0}$.
For
$\{B,K_0,K_\pm ; V_\pm,W_\pm\}$ a basis
of $\ospc$, the above structures are
explicitly displayed in the following commutation
($[\,,\,]$) and anticommutation
($[\,,\,]_+$) relations:   {\renewcommand{\arraycolsep}{26pt}
\[\begin{array}[b]{@{}llr@{}}   \vspace{0.2cm}
\qquad\qquad\qquad\quad[K_0,K_\pm]=\pm K_\pm,
&[K_+,K_-]=-2K_0,
&(2.2\hbox{a})\\
\vspace{0.2cm}
\qquad\qquad\qquad\quad[B,K_\pm]=0,  &[B,K_0]=0,
&(2.2\hbox{b})\\
\vspace{0.2cm}
\qquad\qquad\qquad\quad[K_0,V_\pm]=\pm{1\over2} V_\pm,
 &[K_0,W_\pm]=\pm{1\over2}  W_\pm,  &\qquad
\quad\,(2.2\hbox{c}) \\
\vspace{0.2cm}
\qquad\qquad\qquad\quad[K_\pm,V_\pm]=0, &[K_\pm,W_\pm]=0,
&(2.2\hbox{d}) \\
\vspace{0.2cm}
\qquad\qquad\qquad\quad[K_\pm,V_\mp]=\mp V_\pm,
&[K_\pm,W_\mp]=\mp
W_\pm, &(2.2\hbox{e})\\
\vspace{0.2cm}
\qquad\qquad\qquad\quad[B,V_\pm]={1\over2}  V_\pm,
&[B,W_\pm]=-{1\over2}
W_\pm, &(2.2\hbox{f})\\
\vspace{0.2cm}
\qquad\qquad\qquad\quad[V_\pm,V_\pm]_+ =0,
&[ W_\pm,W_\pm]_+=0,
&(2.2\hbox{g})\\
\vspace{0.2cm}
\qquad\qquad\qquad\quad[V_\pm,V_\mp]_+=0,
&[W_\pm,W_\mp]_+ =0,
&(2.2\hbox{h})\\
\qquad\qquad\qquad\quad[V_\pm,W_\pm]_+ =K_\pm,
 &[V_\pm,W_\mp]_+ =K_0\mp B. &(2.2\hbox{i})\\
\end{array}\]}\par
\addtocounter{equation}{1}

Clearly, the even component is spanned by $\{B,K_0,K_\pm\}$.  The
$K$-generators form an $\ssu$ Lie subalgebra (see (2.2a)),
and $B$ spans a
one dimensional center of  $\ospc_{\bar 0}$ (see (2.2b)).
On the other hand, the
odd component is the span of $V_\pm$ and $W_\pm$.
More precisely,
$\{V_\pm\}$ (resp.
$\{W_\pm\}$) span $\ospc_1$ (resp. $\ospc_{-1}$).
The fact that each of these
two-dimensional vector spaces carry an irreducible
representation of $\ospc_{\bar 0}$ is transparent from
equations (2.2c)--(2.2f);
$\ospc_1$ and $\ospc_{-1}$
are distinguished by the distinct eigenvalues of $B$ in (2.2f).

In order to construct irreducible highest (or lowest)
weight representations of
$\osp$ one needs to exhibit a Borel subsuperalgebra.
In other words, we
look for a decomposition of $\ospc$ of the following form:
\beq
\ospc=\n^-\oplus \hh \oplus\n^+, \label{n}
\eeq
where $\hh$ is a Cartan subalgebra of $\ospc_{\bar 0}$,
and
$\n^-$ and  $\n^+$ are subsuperalgebras of $\ospc$ such that
$[\hh,\n^\pm]\subseteq
\n^\pm$.  The subsuperalgebras $\b^+=\hh\oplus \n^+$
and  $\b^-=\hh\oplus
\n^-$ are called respectively the positive and the negative Borel
subsuperalgebras.  In terms of the basis of $\ospc$ given above
it clearly appears that $\hh$ is the complex span of $\{B,K_0\}$.
Moreover, $\n^+_{\bar 0}$ (resp. $\n^-_{\bar 0}$)
can be taken to be  the
span of $\{K_+\}$ (resp. $\{K_-\}$).  Having fixed this, it can easily
be shown that there exist three possible Borel subsuperalgebras of
$\ospc$.  Since
$\b^\pm=\hh\oplus\n_{\bar 0}^\pm\oplus\n_{\bar 1}^\pm$, we need only
to exhibit the three possible $\n_{\bar 1}^\pm$.  These are:
\begin{description}
\item[(i)\phantom{ii}] $\n_{\bar 1}^+=\hbox{span}\{V_+,V_-\}$ and
$\n_{\bar 1}^+=\hbox{span}\{W_+,W_-\}$,
\smallskip
\item[(ii)\phantom{i}] $\n_{\bar 1}^+=\hbox{span}\{W_+,W_-\}$ and
$\n_{\bar 1}^-=\hbox{span}\{V_+,V_-\}$,
\smallskip
\item[(iii)] $\n_{\bar 1}^+=\hbox{span}\{V_+,W_+\}$ and $\n_{\bar
1}^-=\hbox{span}\{V_-,W_-\}$.
\end{description}
Few remarks are now in order.
\begin{description}
\item[Remark 2.1:]  The situation in {\bf (i)}
is symmetric to that in {\bf (ii)}.  Moreover, they both fit with the
\hbox{$\Z$-grading} given in (\ref{2.1}).
Indeed, we have that $\ospc_0=\n^-_{\bar
0}\oplus\hh\oplus\n^+_{\bar 0}$ and $\ospc_{\pm1}=
\n_{\bar 1}^\pm$.
\item[Remark 2.2:]  The $\pm$ indices carried by the
$V$s and the
$W$s are misleading regarding the root space decomposition
of $\ospc$
relative to the $\hh$ given above.  Indeed, one can see
from (2.2i) that if
$V_+$ (resp. $W_+$) is associated to some odd root
$\alpha$, then $W_-$
(resp. $V_-$) is associated to $-\alpha$.  This
unconventional choice of
notations is aimed at making the $\Z$-gradation
of (\ref{2.1}) explicit  in
the defining relations of $\ospc$ (see (2.2a)--(2.2i)).
\item[Remark 2.3:] Since the set of positive roots
depends on the positive Borel subsuperalgebra
considered, we have then
here three possible sets of this type.  While the set of
positive roots
arising from the case {\bf (iii)} contains two odd simple
roots,
those arising from the cases {\bf (i)} and {\bf (ii)} contain one
odd and one even
simple roots (recall that
rank$(\ospc)\!=\!2$).  The systems of simple roots of both {\bf (i)}
and {\bf (ii)} can be connected to each other by an element of the
Weyl group of
$\ospc$ (${\cal W}(\ospc):={\cal W}(\ospc_{\bar 0})$); they then
give rise
to the same Dynkin diagram.  On the other hand,
the system of simple roots
of {\bf (iii)} does not belong to the ${\cal W}(\ospc)$-orbit
containing the
two previous cases; a different Dynkin diagram arises then.
This situation
is a special feature of basic classical simple Lie superalgebras
\cite{Nino}.  Indeed, usual complex simple Lie algebras admit a
unique Dynkin diagram.  The two Dynkin diagrams mentioned
above exhaust all the possibilities for $\ospc$ \cite{Nino}.
Notice finally
that Kac considered in his original classification \cite{Kac} only
those Borel subsuperalgebras that lead to the minimum
number of odd
simple roots.  They were called {\it distinguished\/} Borel
subsuperalgebras.
\item[Remark 2.4:]  Contrary to the two other cases, the
choice in {\bf (iii)}
leads to a very interesting \hbox{$\Z$-grading} of
$\ospc$
\cite{Nish}.  Both different and finer than that of equation
(\ref{2.1}), it is given by:
\beq
\ospc = {\buildrel(-2)\over{\n^-_{\bar 0}}}
\oplus {\buildrel(-1)\over{\n^-_{\bar 1}}}
\oplus
\!\!{\buildrel{(0)}\over{\phantom{{}^+}\hh\phantom{{}^+}}}\!\!
\oplus {\buildrel(\,1\,)\over{\n^+_{\bar 1}}}
\oplus {\buildrel(\,2\,)\over{\n^+_{\bar 0}}}.
\eeq
\end{description}
The abstract lowest weight representations that will be
described below
are those obtained using the Borel subsuperalgebras of
case {\bf (iii)}.
This choice is justified simply by the fact that, up to
some discrepancies
that we correct here, the associated representations
are those already
used in \cite{Balal2}.  A more mathematical description
of these
representations is given in \cite{Nish}.

An {\it abstract irreducible lowest-weight $\osp$-module\/}
is explicitly
constructed starting from a lowest-weight state.  According to our
choice of Borel subsuperalgebras,  namely the one made above in {\bf
(iii)}, the lowest-weight state is the state, temporarely denoted $\ko$,
which is simultaneously annihilated by
$K_-$,
$V_-$ and $W_-$, and which moreover is a common eigenstate of
both the
Cartan subalgebra generators, $B$ and $K_0$.  Hence, $\ko$
is such that,
\beq {K}_0 \ko =\tau \ko,
\quad {B} \ko =b \ko, \label{2.7}
\eeq and
\beq  {K}_- \ko = {V}_- \ko = {W}_- \ko =0, \label{2.8}
\eeq
where $0<\tau\in\R$ and $b\in\R$ completely
specify a  lowest-weight $\osp$-module, which is
denoted $\vtaub$
throughout.  A basis of $\vtaub$ is explicitly obtained
by applying to
$\ko$  basis elements of the enveloping superalgebra of the
Lie
subsuperalgebra $\n^+$ (as given in (\ref{n}) and {\bf (iii)} above).
The
results of this construction are now displayed; more details are
given in
Appendix~B.

The following observation simplifies the construction.   By
restricting (\ref{2.7}) and (\ref{2.8}) to the $\ssu$-generators,
$K_0$
and $K_-$, it clearly appears that $\ko$ is also the lowest-weight
vector
of an irreducible $\ssu$-module
$\dtau\subset\vtaub$, namely,
\beq
\dtau\equiv\hbox{span}\{|\tau, \tau+m\rangle, m \in \N\}.\label{discrete}
\eeq
This is the  representation space of the well known {\it positive
discrete series representations\/} of $\ssu$.  As a subspace of
$\vtaub$, $\dtau$ is  an eigenspace of
$B$ with eigenvalue $b$. This is an immediate consequence of both
(\ref{2.7}) and the fact that $B$ commutes with $\ssu$ (see (2.2b)).
Combined with (2.7), these facts suggest the following
notation for the lowest-weight state of $\vtaub$, namely
$\ko\equiv\lw$.

As it is explicitly shown in Appendix~B, $\vtaub$ is built out of more
than one irreducible $\ssu$-module.  Two cases must however  be
distinguished, namely, either $|b|<\tau$ or $b=\pm\tau$.

\begin{description}
\item[Typical:]  When $|b|<\tau$, as a vector space,
$\vtaub$, turns out to be the direct sum of irreducible lowest-weight
$\ssu$-modules (positive discrete series). More precisely,
\beq
\vtaub\equiv\dtau\oplus 2\cdot D(\tau+\textstyle\demi)\oplus
D(\tau+1).\label{vtaub1}
\eeq
Here $D(\tau+\demi)$ appears with multiplicity $2$.  These two copies of
$D(\tau+\demi)$ are distinguished eigenspaces of $B$, with eigenvalues
$b+\demi$ and $b-\demi$.   The degeneracy is then raised when one
considers the $\ssu$-modules appearing in (\ref{vtaub1}) as
($\ssu\oplus{\rm so}(2)$)-modules. An extra subscript
has then to be added to our previous notation in order to take
this fact into
account.  Observing that $D(\tau+1)$, as $\dtau$, is a
$B$-eigenspace with eigenvalue
$b$, we can write the following finer decomposition of $\vtaub$,
\beq
\vtaub=D_b(\tau)\oplus D_{b+\demi}(\tau+\textstyle\demi)\oplus
D_{b-\demi}(\tau+\textstyle\demi)\oplus D_b(\tau+1).\label{vtaub2}
\eeq
More precisely,
\begin{eqnarray}
\vtaub &\!\!\!\equiv\!\!\! &\hbox{span}\{|b, \tau, \tau+m\rangle,
|b+\textstyle\demi,
\tau+\textstyle\demi, \tau+\textstyle\demi+m\rangle,
 \nonumber\\ &\!\!\!\!\!\! & \qquad |b-\textstyle\demi,
\tau+\textstyle\demi, \tau+\textstyle\demi+m\rangle,
|b,\tau+1, \tau+1+m\rangle; m \in \N\}. \label{vtaub3}
\end{eqnarray}
Moreover, as a vector superspace, $\vtaub=V_{\bar 0}(\tau,b)\oplus
V_{\bar 1}(\tau, b)$, where $V_{\bar 0}(\tau,b)\equiv
D_b(\tau)\oplus D_b(\tau+1)$ and $V_{\bar 1}(\tau,b)\equiv
D_{b+\demi}(\tau+\textstyle\demi)\oplus
D_{b-\demi}(\tau+\textstyle\demi)$.
(The action of the generators of $\ospc$ in $\vtaub$ is
explicitly given in
Appendix~B.)  The irreducible
$\osp$-modules, $\vtaub$, for $|b|<\tau$ are usually called
{\it typical
representations\/}
\cite{Cornwell,Scheunert,Kac,ShNaRitt}.
\medskip
\item[Atypical:]  When $b=\tau$ (resp. $b=-\tau$),
$V(\tau, \tau)$ (resp.  $V(\tau, -\tau)$) is no longer irreducible.
It contains an
$\osp$-submodule,  $V'(\tau,
\tau)\equiv D_{\tau+\demi}(\tau+\demi)\oplus D_\tau(\tau+1)$
(resp.
$V'(\tau, -\tau)\equiv D_{-\tau-\demi}(\tau+\demi)\oplus
D_{-\tau}(\tau+1)$), generated by the {\it primitive vector\/} $V_+\ko$
(resp.
$W_+\ko$).  The quotient
$V(\tau, \tau)/V'(\tau, \tau)$ (resp. $V(\tau, -\tau)/V'(\tau, -\tau)$)
appears then as the appropriate irreducible $\osp$-module.  More
precisely,
\beq U(\pm\tau)\equiv V(\tau,\pm\tau)/V'(\tau, \pm\tau)\equiv
D_{\pm\tau}(\tau)\oplus D_{\pm(\tau-\demi)}(\tau+\textstyle\demi).
\label{vtaub4}
\eeq
These representations are known as the {\it atypical
representations\/} \cite{Cornwell,Scheunert,Kac,ShNaRitt}.
\end{description}

\noindent The above typical
and atypical $\osp$-modules can be turned into {\it super unitary\/}
irreducible representations \cite{Nish}, by equipping $\vtaub$ with a  {\it
super Hermitian\/} form $\langle\, \cdot\, |\, \cdot\,\rangle$.  The
latter  is a notion originally defined in \cite{SternWolf}, and used in the
context of representation theory of Lie superalgebras in \cite{Scheunert},
and more recently  in \cite{Nish}.  It is defined in the  following way:
\beq
\langle\, \cdot\, | \,\cdot\,\rangle:
\vtaub\times\vtaub\longrightarrow\C   \label{SH1}
\eeq
such that $\forall\ u, v$ two homogeneous elements of $\vtaub$
\beq
\overline{\langle u | v\rangle}=(-1)^{\epsilon(u)\epsilon(v)}\langle v |
u\rangle,\label{SH2}
\eeq
where the parity of a homogeneous element $w\in\vtaub$ is
$\epsilon(w)\equiv 0(1)$ for $w \in  V_{\bar 0(\bar 1)}(\tau, b)$.   The
elements of $V_{\bar 0(\bar 1)}$ are called even (odd) elements of
$\vtaub$.  The super Hermitian form is taken linear in the
second component. In what follows we will consider a
{\it homogeneous\/} realization of (\ref{SH2}), namely,
$\forall\ u=u_0+u_1,\ v=v_0+v_1 \in
\vtaub=V_{\bar 0}(\tau, b)\oplus V_{\bar 1}(\tau, b)$,
\beq
\langle u | v\rangle\equiv\langle u_0 | v_0\rangle_0+i\langle u_1 |
v_1\rangle_1,\label{SH2'}
\eeq
where, $\langle\, \cdot\, | \,\cdot\,\rangle_0$ (resp.  $\langle\,
\cdot\, |
\,\cdot\,\rangle_1$) is an Hermitian form, in the usual sense, on
$V_{\bar 0}(\tau, b)$ (resp. $V_{\bar 1}(\tau, b))$.
Since $\overline{\langle u_{k} | v_{k}\rangle}_{k}=\langle v_{k} |
u_{k}\rangle_{k}$ for $k=0,1$, one clearly sees that (\ref{SH2'})
satisfies
(\ref{SH2}).  The super Hermitian form in (\ref{SH2'}) is even.

Besides being an $\osp$-module,
$\vtaub$ will be also considered as a left $\cal B$-module, where
${\cal B}={\cal B}_0\oplus {\cal B}_1$ is a {\it complex Grassmann
algebra\/} \cite{Berezin3,Cornwell}.   The super  Hermitian form
(\ref{SH1})--(\ref{SH2'}) will then be extended to the {\it Grassmann
envelope of the second kind\/}
\cite{Berezin3}
$\tvtaub$ of
$\vtaub$,
\beq
\tvtaub\equiv\left({\cal
B}\otimes\vtaub\right)_0,\label{tvtaub}
\eeq
such that
\beq
\langle\, \cdot\, | \, \cdot\,\rangle:
\tvtaub\times\tvtaub\longrightarrow{\cal B}_0\label{SH3}
\eeq
{}From the next section on, ${\cal B}$ will assume
a very specific form, as the  exterior algebra over $\C^4$, namely, ${\cal
B}\equiv\bigwedge\!{\C}^4$ (see Appendix~A for more details about
(\ref{tvtaub}) and (\ref{SH3})).

\begin{description}
\item[Remark 2.5:]  The super Hermitian form in
(\ref{SH1}) and (\ref{SH2'}) is {\it positive definite\/} on $\vtaub$ for
$|b|<\tau$, in the sense that the Hermitian forms on both $V_{\bar
0}(\tau, b)$ and $V_{\bar 1}(\tau, b)$ are positive definite.  In
Appendix~B,
we show how this structure turns the  typical module $\vtaub$ into
a super unitary irreducible representation of \osp.
\smallskip
\item[Remark 2.6:]  The construction of atypical representations
(\ref{vtaub4}) looks very much like the construction of the so-called
indecomposable representations, which usually describe massless
relativistic quantum elementary systems.  This analogy is confirmed
by the fact that the super Hermitian form (\ref{SH2'}) is
no longer positive definite on $V(\tau, \pm\tau)$.  Indeed, as it is
shown
in Appendix~B, both $V'(\tau,
\tau)$ and
$V'(\tau, -\tau)$ are $\osp$-submodules of {\it zero-norm\/} states,
with respect to the super Hermitian form.  However, notice that the
atypical representations are,  as the typical ones, simply expressed in
terms of discrete series representations of $\ssu$ (see (\ref{vtaub4})).
\smallskip
\item[Remark 2.7:]  It is important to note that the
atypical modules $U(\pm\tau)$ in (\ref{vtaub4}) are irreducible
osp$(1/2)$-modules, where osp$(1/2)$ stands here for the Lie
subsuperalgebra of $\osp$ whose Cartan-Weyl basis is
$\{K_0, K_\pm, {1\over \sqrt2}(V_\pm+W_\pm)\}$.   This interesting
observation will be
discussed further on in Section~6.
\end{description}

\noindent We  now briefly discuss Schur's lemma.  Since $\ospc$ is
of rank $2$,  the center of its enveloping superalgebra
is generated by two
$\ospc$-invariants, which are, respectively, quadratic
(the Casimir) and
cubic in the generators of $\ospc$.  For simplicity, we exhibit below
only the explicit expression of the former, which we denote $Q_2$.
Hence,
\beq  Q_2=C_2-B^2+K_0-W_+V_--V_+W_-,
\eeq  where $C_2$ is the Casimir invariant of the $\ssu$ subalgebra,
namely,
\beq C_2=K_0^2-K_0- K_+K_-.\label{C2}
\eeq The irreducibility of $\vtaub$ implies that, on $\vtaub$, both
invariants are constant multiples of the identity.  The exact value of
the quadratic constant on
$\vtaub$ is simply obtained by evaluating $Q_2$ on the lowest-weight
state $\lw$.  Hence,
\beq
Q_2\equiv (\tau^2-b^2)\I\quad \hbox{on\quad $\vtaub$,\quad since}
\quad C_2\equiv
\tau(\tau-1)\I
\quad\hbox{on\quad $\dtau\subset\vtaub$}.
\eeq
One clearly sees that $Q_2$ is identically zero on the atypical modules.

Finally, a lowest-weight irreducible $\osp$-module is said to
be an {\it integrable module\/}, whenever it is also a module for the
{\it double covering\/} of the body Sp$(2,\R)\times$SO$(2)$ of $\OSp$.
Since  the maximal torus in $\OSp$  (and in Sp$(2,\R)\times$SO$(2)$ too)
is the $U(1)\times U(1)$ subgroup generated by $K_0$ and
$B$, one easily sees from (\ref{2.7}) that
$\vtaub$ is an integrable $\osp$-module if and only
if both $\tau$ and $b$ are
half integers.   Throughout we will only deal with
integrable super unitary
irreducible $\osp$-modules, and we will no
longer mention this fact.  More precisely, we
will restrict our attention to the typical integrable
representations, knowing that
all the constructions of the forthcoming
sections carry over to the
atypical case.  However, some very interesting
points concerning the latter are
worth to be mentioned.  They are gathered in Section~6.

\subsection{OSp(2/2) coherent states}

In this section we construct the $\OSp$-coherent states
associated to the typical representations described in the previous
section.   We
should mention that, up to some discrepancies, that are cured here, and
a different choice of conventions (see Appendix~A),
this construction was
originally carried out in \cite{Balal2}.

As stressed in the introduction,  the coherent states for a Lie group
$G$ are the quantum states belonging to the orbit of $G$
through a fiducial
vector in an irreducible $\g$-module, carrying a unitary representation of
$G$.  Here, $\g$ stands for the Lie algebra of $G$.
Recall that we also
mentioned in the introduction that the minimal requirement for the
construction of $G$-CS consists in an irreducible unitary $\g$-module.
For $\OSp$, irreducible super unitary $\osp$-modules are
at our disposal.

Considering the lowest-weight state $\lw\in \vtaub$ as the fiducial state,
taking into account equations (\ref{2.7})--(\ref{2.8}), and extending in a
straightforward manner Perelomov's construction \cite{Pere}, the
typical $\OSp$-coherent states, are obtained as follows:
\beq
|a, \t,\chi\rangle\equiv {\cal N} \exp(a {K}_+ +\t {V}_+ +
\chi {W}_+)\lw.\label{atx}
\eeq
They belong to $\tvtaub$ (see (\ref{vtaub3}) and (\ref{tvtaub})), for ${\cal
B}$ the Grassmann algebra generated by the
complex anticommuting variables $\t$,
$\chi$, and their complex conjugates, $\tb$ and $\chib$.
These are odd elements
of ${\cal B}$, while $a$ is an even element of $\cal B$.
More details about $\cal B$
are given in Appendix~A.  In (\ref{atx}),
${\cal N}$ is a normalization factor
which will be explicitly determined below.

Since,
$[{K}_{+},{V}_{+}]=0$, $[{K}_{+},  {W}_{+}]=0$ and
$\t^2=\chi^2=0$, we can rewrite (\ref{atx}) as follows,
\beq
|a, \t,\chi\rangle= {\cal N} [\exp(a {K}_+)][1+\t{V}_+ +\chi{W}_+
+\textstyle{1\over2}\chi\t({V}_+ {W}_+ - {W}_+ {V}_+)]\lw .
\eeq
A simple computation, based on (\ref{B15})--(\ref{Action}), leads to:
\begin{eqnarray} |a, \t,\chi\rangle &\!\!\!=\!\!\!& {\cal N}
\biggl[\exp\biggl(\biggl(a +
\textstyle{b\over 2\tau} \chi\t\biggr) K_+\biggr)\biggr] \biggl(
|b,\tau,\tau\rangle+\t\sqrt{\tau-b}\,
|b+\textstyle{1\over2},\tau+\textstyle{1\over2},\tau+
\textstyle{1\over2}\rangle  \label{2.14}  \\ &\!\!\!\!\!\!& \mbox{} + \chi
\sqrt{\tau+b} \, |b-\textstyle{1\over2},\tau+\textstyle{1\over2},
\tau+\textstyle{1\over2}\rangle +\chi\t \sqrt{
\textstyle{(\tau^2-b^2)(2\tau+1)\over 2\tau}}\, |b,\tau+1,\tau+1\rangle
\biggr)\! . \nonumber
\end{eqnarray}
At this point, we introduce the new variable
\beq
z=a +{b\over 2\tau}\chi\t \in \C.\label{z}
\eeq
Notice that for simplicity we are choosing here (and throughout)
$z$ to be a
complex number.  This choice is discussed in
Section~7.  Finally, using the known action of
${K}_+$ in
$\vtaub$ (\ref{B18}), we get the explicit form of the typical coherent
states (\ref{2.14}):
\begin{eqnarray} |z, \t,\chi\rangle & = & {\cal N}\left[\,
\sum_{m=0}^{\infty}
\sqrt{\textstyle{\Gamma(2\tau+m)\over m!\, \Gamma(2\tau)}}\ z^m
|b,\tau,\tau+m\rangle   \right. \nonumber  \\ &  &\mbox{} + \t \sqrt{\tau-b}
\sum_{m=0}^{\infty}
\sqrt{\textstyle{\Gamma(2\tau+m+1)\over m!\, \Gamma(2\tau+1)}}\ z^m
|b+\textstyle{1\over2},\tau+\textstyle{1\over2},\tau+\textstyle{1\over2}+
m\rangle    \label{cs} \\ &  &\mbox{} + \chi \sqrt{\tau+b}
\sum_{m=0}^{\infty}
\sqrt{\textstyle{\Gamma(2\tau+m+1)\over m!\, \Gamma(2\tau+1)}}\ z^m
|b-\textstyle{1\over2},\tau+\textstyle{1\over2},\tau+\textstyle{1\over2}+
m\rangle  \nonumber \\  &  &\left. \mbox{} + \chi\t
\sqrt{\textstyle{(\tau^2-b^2)(2\tau+1)\over 2\tau}}\,
\sum_{m=0}^{\infty}
\sqrt{\textstyle{\Gamma(2\tau+m+2)\over m!\, \Gamma(2\tau+2)}}\ z^m
|b,\tau+1,\tau+1+m\rangle  \right].  \nonumber
\end{eqnarray}
Now, using the super Hermitian form (\ref{SH3}), the fact that
the basis of $\vtaub$ in (\ref{vtaub3}) is super-orthonormal with
respect to (\ref{SH2'}) (see Appendix~B), and the conventions of
Appendix~A, one
easily evaluates
${\cal N}$. Hence,
\beq
{\cal N}=(\de)^{\tau}\left[ 1+i{\tau-b\over 2}{\tb\t\over \de} +
i{\tau+b\over 2} {\chib\chi\over \de} - {(\tau^2-b^2)(\tau-1)\over 4\tau}
{\tb\chib\chi\t\over (\de)^2}
\right]. \label{N}
\eeq
Notice here that for obvious reasons we have considered ${\cal
N}$ real.  Moreover, the following identity,
\beq
{1\over (1-x)^s}= \sum_{m=0}^{\infty}  {\Gamma(s+m)\over m!\,
\Gamma(s)}\ x^m,\quad \hbox{for}\quad |x|<1 \label{ID},
\eeq
has been used in order to write the result (\ref{N}) in a compact form.
It is
worth mentioning at this point that when one sets to zero the odd
variables $\t$ and $\chi$, all the previous formulae reduce
exactly to those of the SU$(1,1)$-CS \cite{Pere}.  In that case, and
clearly here too, $z$ spans the unit disc $\du=\{z \in \C\ ; \ |z|<1\}$;
$\du\equiv\su/$U$(1)$.

For a later use, we evaluate now the $|\zb, \tb,
\chib\rangle$ CS, where,
$\zb$, $\tb$ and $\chib$ are the complex conjugates of $z$,
$\t$ and $\chi$.  Straightforwardly,
\begin{eqnarray}  |\vb, \tb,\chib\rangle & = & {\cal N}'\left[\,
\sum_{m=0}^{\infty}
\sqrt{\textstyle{\Gamma(2\tau+m)\over m!\, \Gamma(2\tau)}}\
\vb^m  |b,\tau,\tau+m\rangle   \right. \nonumber \\  &  &\mbox{} + \tb
\sqrt{\tau-b}
\sum_{m=0}^{\infty}
\sqrt{\textstyle{\Gamma(2\tau+m+1)\over m!\, \Gamma(2\tau+1)}}\
\vb^m |b+\textstyle{1\over2},\tau+\textstyle{1\over2},\tau+
\textstyle{1\over2}+m\rangle  \label{csb} \\ &  & \mbox{} + \chib
\sqrt{\tau+b}  \sum_{m=0}^{\infty}
\sqrt{\textstyle{\Gamma(2\tau+m+1)\over m!\, \Gamma(2\tau+1)}}\
\vb^m |b-\textstyle{1\over2},\tau+\textstyle{1\over2},\tau+
\textstyle{1\over2}+m\rangle  \nonumber \\ &  &  \left. \mbox{} + \chib
\tb\sqrt{\textstyle{(\tau^2-b^2)(2\tau+1)\over 2\tau}}\,
\sum_{m=0}^{\infty}
\sqrt{\textstyle{\Gamma(2\tau+m+2)\over m!\, \Gamma(2\tau+2)}}\
\vb^m |b,\tau+1,\tau+1+m\rangle  \right], \nonumber
\end{eqnarray}
the normalizing constant being now
\beq {\cal N}'=(\de)^{\tau}\left[ 1-i {\tau-b\over 2} {\tb\t\over \de} -i
{\tau+b\over 2} {\chib\chi\over \de} - {(\tau^2-b^2)(\tau-1)\over 4\tau}
{\tb\chib\chi\t\over (\de)^2}
\right].\label{N'}
\eeq

Having the explicit form of the typical $\OSp$-CS, we can now apply to
them a super extension of Onofri's analysis \cite{Onofri} in order to reveal
the (super)geometry underlying them, or equivalently, in order to evaluate
the classical limit of the quantum theory described by the typical
representation $\vtaub$ of the previous subsection.  This  analysis is
the main concern of the next two sections.

%%%%%%%%%%%%%%%%%%%%%%%%%%%%%%%%%%%%%%
%%%% 3. N=2 superunit disc is supersymplectic %%%%%
%%%%%%%%%%%%%%%%%%%%%%%%%%%%%%%%%%%%%%
\section{Supersymplectic geometry  and the OSp(2/2)-CS}\label{sec3}

\noindent In analogy with the non-super case \cite{Pere}, the results
of the previous section suggest that the space parametrizing the typical
$\OSp$-CS is the $\OSp$-homogeneous superspace $\OSp/\left(
U(1)\times U(1)\right)$, realized in terms of the coordinates
$(z,\t,\chi)$, where $z$ parametrizes the unit disc
$\du:=\{z\in\C;\, |z|<1\}$. This
realization will be subsequently called the $N=2$ super unit disc, and it
will be denoted $\dud$.  We recall that the $N=1$ super unit disc $\duu$
was fully considered in
\cite{Amine1,Amine2}.  We will consider here $\dud$ as a supermanifold,
although its complete and precise geometric characterization will be only
given in Section~4.

In Section~3.1, we carry on super extending Onofri's analysis \cite{Onofri},
which only makes use of the explicit form of the CS.  This analysis will
provide us with a partial description of the geometric structure underlying
the typical $\OSp$-CS.  Some of the results obtained here will be
revisited in Section~4, in the light of the general theory of
supermanifolds.   Let us recall that in
the case of a semi-simple Lie group, starting from the associated CS,
Onofri's analysis allows one to equip the homogeneous space
parametrizing
these CS with an invariant symplectic form (i.e. a closed and
non-degenerate
$2$-form), which is moreover K\"ahler.   Analogously, $\dud$ is
equipped here with an invariant {\it supersymplectic\/} form, in the sense
of
\cite{Berezin3,Kost2,Roth1}.  Moreover, as it will be discussed in Section
4, this form is {\it super K\"ahler\/} in a sense to be defined (see
also \cite{Amine1,Amine2}). In Section~3.2, extending Berezin's notion of
covariant symbols, we evaluate the classical observables (superfunctions
on
$\dud$) associated to the infinitesimal action of $\OSp$ on $\dud$.  By
doing this we exhibit a moment map that identifies $\dud$ with an
$\OSp$-coadjoint orbit.  In the process, the Hamiltonian vector
superfields associated to the classical observables are computed.
Finally, in Section~3.3 we evaluate the {\it  Liouville super measure\/}
on $\dud$.  This will be needed in Section~5.

\subsection{The supersymplectic form}

Onofri's analysis starts by evaluating a real function from the $G$-CS; it
plays the role of
a K\"ahler potential for a $G$-invariant K\"ahler form on the space
parametrizing the $G$-CS.  Here, emphasizing the symplectic output of
this procedure, we extend it to our typical $\OSp$-CS.  Hence, from
equation (\ref{csb}), we evaluate the following superfunction on ${\cal
D}^{(1\vert 2)}$,
\begin{eqnarray}
f(z,\vb,\t,\tb,\chi,\chib) &\!\!\!:=\!\!\!& \log |\langle 0\ketb|^{-2}
\nonumber\\ &\!\!\!=\!\!\!& 2\tau\left[-\log (\de)+ i {\tau-b\over 2\tau}
{\tb\t \over
\de}\right.
\nonumber \\
& & \left.\phantom{-\log (\de)}+ i {\tau+b\over 2\tau} {\chib\chi\over
\de} - {\tau^2-b^2\over 4\tau^2} {\tb\chib\chi\t\over (\de)^2}
\right]. \label{pot}
\end{eqnarray}
Let now $d$ be the exterior derivative on $\dud$ given by $d=\delta+{\bar
\delta}$, where
\beq
\delta=dz\pv +d\t\pt+d\chi\pchi,\qquad  {\bar \delta}=d\vb\pvb
+d\tb\ptb+d\chib\pchib.
\eeq
Notice that here $\epsilon(d)=0$, i.e. $d$ is an even quantity.

An even two-superform on ${\cal D}^{(1\vert 2)}$ can now be
obtained from
the  superfunction in (\ref{pot}) in the following way:
\beq
\omega\equiv-i \delta{\bar \delta} f=\omega_0 +\omega_2+\omega_4,
\label{sform1}
\eeq
where,
\begin{eqnarray}
\omega_0  &\!\!\!\!=\!\!\!\! & {-2i \tau\over (\de)^2}\,dz\,\!d\vb; \label{O0}
\\
\omega_2 &\!\!\!\!=\!\!\!\!&  -(\tau-b)  \left[{1\over (\de)}\,d\t\,\!d\tb +
{1\over (\de)^2} [\t\vb\, dz\,\!d\tb-\tb z\, d\t\,\!d\vb] - \tb\t\,
{(1+|z|^2)\over (\de)^3}\,dz\,\!d\vb
\right] \nonumber\\  &\!\!\!\!\!\!&\mbox{} - (\tau+b)\left[  {1\over
(\de)}\,d\chi\,\!d\chib + {1\over (\de)^2} [\chi\vb\, dz\,\!d\chib-\chib z\,
d\chi\,\!d\vb ] \right.
\nonumber\\  &\!\!\!\!\!\!& \mbox{} \phantom{{1\over (\de)^2} [\chi\vb\,dz
d\chib-\chib z\, d\chi d\vb ][\chi\vb\,dz -\vb\,dv
d\chib]}\left.-\chib\chi\,{(1+|z|^2)\over (\de)^3}\, dz\,\!d\vb
\right];  \label{O2} \\
\omega_4 &\!\!\!\!=\!\!\!\!& -i {\tau^2-b^2\over 2\tau(\de)^2}
\left[-2 {(1+2|z|^2)\over (\de)^2}\,\tb\chib\chi\t\, dz\,\!d\vb+
\tb\t\, d\chi\,\!d\chib +\chib\chi\, d\t\,\!d\tb -\tb\chi\, d\t\,\!d\chib
\right.
\nonumber \\   &\!\!\!\!\!\!& \mbox{}\left. -\chib\t\, d\chi\,\!d\tb +
{2\,\chib\chi\over (\de)}  [\t\vb\,  dz\,\!d\tb-\tb z\,  d\t\,\!d\vb]  +
{2\,\tb\t\over (\de)} [\chi\vb\,  dz\,\!d\chib-\chib z\,  d\chi\,\!d\vb]
\right]\!\!.  \label{sform2}
\end{eqnarray}
This two-superform belongs to the exterior algebra on $\dud$.  The
latter is a bi-graded $\Z\times\Z_2$ algebra \cite{Kost2}, where the
$\Z$-gradation is the usual gradation of de Rham complexes, while the
$\Z_2$-gradation is the natural gradation accompanying supersymmetry
(i.e. the $\Z_2$-gradation of the Grassmann algebra ${\cal B}$).   More
precisely, for any two superforms
$\beta_1$ and $\beta_2$ on
$\dud$, one has:
\beq
\beta_1\beta_2=(-1)^{a_1a_2+b_1b_2}\beta_2\beta_1,
\eeq
where
$a_i$ (resp. $b_i$) is the degree of the superform $\beta_i$ with respect to
the $\Z$ (resp. $\Z_2$) gradation.  Hence, in (\ref{sform2}), $dz\,
d\zb=-d\zb\,dz$ (this is the usual wedge product),
$dz\,d\tb=-d\tb\,dz$ and
$d\t\,d\tb=d\tb\,d\t$.

Using these conventions one can  check by  explicit
calculations that $\omega$ is closed, i.e.
$d\omega=0$. In fact, $\omega$ is closed by construction.  This is a direct
consequence of (\ref{sform1}).  Indeed, since $d=\delta+\bar\delta$,
$d^2=0$ implies that
$\delta^2=\bar\delta^2=\delta\bar\delta+\bar\delta\delta=0$. Hence,
$(\dud,\omega)$ is a supersymplectic supermanifold.  This particular
point will be analyzed further on in Section~4.

\subsection{The classical observables}

\noindent By analogy with the non-super case, one can determine the
classical observables associated to the generators of the supersymplectic
action of $\OSp$ on $\dud$.  This is achieved through the evaluation
of the so-called Berezin covariant symbols \cite{Berezin1}.
As for the superfunction (\ref{pot}), these are obtained simply from the
knowledge of the explicit form of the $\OSp$-CS and the representation
$\vtaub$ they belong to.  Hence, the classical observable $H^{cl}$
associated
to an $\OSp$-generator $H\in\{ {B},{K}_0,{K}_{\pm},{V}_{\pm}, {W}_{\pm}\}$
is given by the Berezin symbol:
\beq
H^{ cl}\equiv\ H^{ cl}(z,\vb,\t,\tb,\chi,\chib):=
\langle\vb,\tb,\chib|{H}\ketb. \label{bsymb}
\eeq
After lengthy but straightforward computations, based on (\ref{csb})
and
results from Appendix~B, one obtains:

\begin{eqnarray}
B^{ cl}&\!\!\!=\!\!\!\!&\displaystyle b+ i {\tau-b\over 2}
{\tb\t\over
\de}- i {\tau+b\over 2} {\chib\chi\over \de},\nonumber \\
K^{cl}_0 &\!\!\!\!=\!\!\!\!& \displaystyle \tau {1+|z|^2\over \de}\left[ 1+ i
 {\tau-b\over 2\tau} {\tb\t\over (\de)} + i {\tau+b\over 2\tau}
{\chib\chi\over (\de)} - {\tau^2-b^2\over 2\tau^2} {\tb\chib\chi\t\over
(\de)^2}\right], \nonumber \\
K^{ cl}_+ &\!\!\!\!=\!\!\!\!&
\displaystyle{2\tau z\over
\de}\left[ 1+ i
 {\tau-b\over 2\tau} {\tb\t\over (\de)} + i {\tau+b\over 2\tau}
{\chib\chi\over (\de)} - {\tau^2-b^2\over 2\tau^2} {\tb\chib\chi\t\over
(\de)^2}\right],\nonumber \\
K^{ cl}_- &\!\!\!\!= \!\!\!\!& \displaystyle
{2\tau\vb\over \de}\left[ 1+ i
 {\tau-b\over 2\tau} {\tb\t\over (\de)} + i {\tau+b\over 2\tau}
{\chib\chi\over (\de)} - {\tau^2-b^2\over 2\tau^2} {\tb\chib\chi\t\over
(\de)^2}\right], \nonumber\\
V^{ cl}_+ &\!\!\!\!=\!\!\!\!&\displaystyle  {1\over
\de}\left[- i(\tau-b)\t+(\tau+b)z\chib - {\tau^2-b^2\over 2\tau}
 {(i z\tb+\chi)\chib\t\over (\de)}\right], \label{clobs} \\
W^{ cl}_-
&\!\!\!\!= \!\!\!\!& \displaystyle {1\over
\de}\left[(\tau-b)\tb-i(\tau+b)\vb\chi - {\tau^2-b^2\over 2\tau}
 {(\vb\t+i\chib)\tb\chi\over (\de)}\right],\nonumber \\
V^{ cl}_-
&\!\!\!\!=\!\!\!\!&
\displaystyle {1\over
\de}\left[-i(\tau-b)\vb\t+(\tau+b)\chib - {\tau^2-b^2\over 2\tau}
 {(\vb\chi+i\tb)\chib\t\over (\de)}\right],\nonumber \\
W^{ cl}_+
&\!\!\!\!=\!\!\!\!& \displaystyle {1\over
\de}\left[(\tau-b)z\tb-i(\tau+b)\chi - {\tau^2-b^2\over 2\tau}
 {(iz\chib+\t)\tb\chi\over (\de)}\right]. \nonumber
\end{eqnarray}

\noindent The obtained classical  observables satisfy the
following relations:
\beq
\overline{B^{ cl}}= B^{ cl},\quad
 \overline{K^{ cl}_0}= K^{ cl}_0,
\quad \overline{K^{cl}_+}= K^{ cl}_-,\quad
\overline{V^{ cl}_+}= i W^{cl}_-,
\quad
\overline{V^{ cl}_-}= i W^{ cl}_+.\label{reality}
\eeq
\begin{description}
\item[Remark 3.1:] Notice that the Berezin symbols
(\ref{bsymb}) are defined in terms of $\ketb$ instead of $\ket$.
In order to
justify this choice we need to anticipate on future results.
In fact, as it
will be shown in Section~5, this choice leads to the classical
theory, the
quantization of which gives rise to a superholomorphic
representation.  The
anti-superholomorphic one arises as the quantization of the
classical theory obtained from the CS $\ket$.
\end{description}

\noindent The Hamiltonian vector superfield $X_H$ associated to a
classical observable  $H^{cl}$, is the solution of the following defining
equation \cite{Kost2},
\beq
X_H\pint\omega=dH^{cl},
\eeq
where `$\pint$' stands for the inner product and $\omega$ is the
supersymplectic form (\ref{sform1}).  Here we display the
Hamiltonian vector
fields associated to the above observables.  A long
computation leads to the
following:

\begin{eqnarray}
X_B &\!\!\!\!= \!\!\!\!& \displaystyle i {\t\over 2 }\pt - i
{\tb\over 2} \ptb - i
{\chi\over2}
\pchi +  i {\chib\over 2} \pchib, \nonumber\\
X_{K_0}
&\!\!\!\!=  \!\!\!\!& \displaystyle  iz\pv-i\vb\pvb+i {\t\over 2} \pt -
i {\tb\over 2} \ptb + i
{\chi\over 2} \pchi -  i {\chib\over 2} \pchib, \nonumber \\
X_{K_+} &\!\!\!\!=  \!\!\!\!&  \displaystyle
iz^2\pv-i\pvb+iz\t\pt+iz\chi\pchi,\nonumber \\
X_{K_-} &\!\!\!\!=  \!\!\!\!& \displaystyle
i\pv-i\vb^2\pvb-i\vb\tb\ptb-i\vb\chib\pchib, \nonumber \\
X_{V_+}
&\!\!\!\!= \!\!\!\!& \displaystyle \left( {\tau-b\over 2\tau}\right)
z\t\pv -  i\left(
{\tau+b\over 2\tau}\right) \chib\pvb +i\ptb - \left(z+ \left(
{\tau-b\over
2\tau}\right)
\chi\t\right) \pchi, \label{vectors}\\
X_{W_-} &\!\!\!\!=  \!\!\!\!&
\displaystyle \left( {\tau+b\over 2\tau}\right)
\chi\pv - i\left(
{\tau-b\over 2\tau}\right) \vb\tb\pvb -\pt +i\left(\vb-
\left(  {\tau-b\over 2\tau}\right)\tb \chib\right) \pchib,
\nonumber\\
X_{V_-} &\!\!\!\!=  \!\!\!\!&  \displaystyle \left(
{\tau-b\over 2\tau}\right) \t\pv - i
\left( {\tau+b\over 2\tau}\right) \vb\chib\pvb
+i\left(\vb +
\left({\tau+b\over 2\tau}\right) \tb\chib\right) \ptb
-\pchi,\nonumber \\
X_{W_+} &\!\!\!\!=  \!\!\!\!& \displaystyle \left( {\tau+b\over
2\tau}\right) z\chi\pv - i\left( {\tau-b\over 2\tau}\right)
\tb\pvb  -\left(z-
\left({\tau+b\over 2\tau}\right) \chi\t\right) \pt+i\pchib.\nonumber
\end{eqnarray}

\noindent The above vector fields verify:
\beq
\overline{X}_B= X_B,\quad \overline{X}_{K_0}= X_{K_0},
\quad \overline{X}_{K_+}= X_{K_-},\quad
\overline{X}_{V_+} = i X_{W_-}, \quad
\overline{X}_{V_-}= i X_{W_+}.\label{reality'}
\eeq

The supersymplectic structure $\omega$
(\ref{sform1}) on $\dud$ defines a Poisson super
bracket structure, $\{\,
,\,\}$, in the space of smooth superfunctions on
$\dud$, turning it into
a Poisson superalgebra.  Indeed, given any two
smooth superfunctions $g$ and $h$ on $\dud$,
\beq
\{g, h\}\equiv -iX_g\pint dh.
\eeq
A simple computation shows that the classical
observables in
(\ref{clobs}) form a Poisson subsuperalgebra
isomorphic to $\osp$.
Hence, the classical observables provide one with a
 supersymplectic
realization of $\osp$.  What precedes is equivalent
to  say that by evaluating the classical observables in
(\ref{clobs}) we
have in fact exhibited an infinitesimally equivariant
momentum map
\cite{Souriau},
$J:
\dud
\rightarrow
\osp^*\equiv\osp$, that identifies
$\dud$ with an $\OSp$-coadjoint orbit \cite{Kost2}.
The latter is
realized as a $(2|4)$-dimensional subsupermanifold of
$\R^{(4|4)}$,
defined by two constraints, which correspond to the
quadratic and the
cubic
$\osp$-invariants.  Indeed, when the latter are evaluated
in the
supersymplectic realization of $\osp$, they are both identically
constant.
For instance, $Q_2^{cl}\equiv \tau^2-b^2$, (recall here that
$|b|<\tau$).

This completes the first stage of our description of the geometry
underlying the typical $\OSp$-CS.  This turns out to be a
supersymplectic
geometry.  The main purpose of the next and final stage
(Section~4) is
to situate the results of the present section within the
already existing
theory of supersymplectic supermanifolds.  Moreover,
by analogy with the non super case \cite{Onofri}, one is
tempted to
go one step further and  consider $(\dud,\omega)$ as an
example of a
general notion of {\it super K\"ahler supermanifolds\/}, which
has not so
far been seriously studied (see however \cite{Amine1,Amine2}).  In
Section~4, this notion will be given a legitimacy, which will open the
door to super extending the full geometric quantization to the super
K\"ahler context  (see Section~5).

Before carrying out this program, we display now a computation,  the
result of which will be needed later on in Section~5.

\subsection{The Liouville supermeasure}

\noindent An super measure on the super unit disc
$\dud$, which super extends the notion of a Liouville measure on a
symplectic manifold, can now be evaluated starting from $\omega$
(\ref{sform1}) and using Berezin's notion of a density \cite{Manin}.
Up to a
multiplicative constant, an $\OSp$-invariant measure on $\dud$ is
given by:
\beq
d\mu (z,\vb, \t,\tb,\chi,\chib)= {i\over \pi}\,{\rm sdet}
||\omega_{A\bar B}||\ dz\, d\vb\, d\t\, d \tb\, d\chi\, d\chib\,
\eeq
where `sdet'  stands for the superdeterminant (or Berezenian)
\cite{Berezin3}, while
$||\omega_{A\bar B}||$ stands for the supermatrix form of $\omega$,
namely,
\begin{eqnarray}
\omega &\equiv & dx^A\,\omega_{A\bar B} \, dx^{\bar B}\nonumber \\
 &= &dz  (a)d\vb +dz  (\alpha)d\tb +dz (\beta) d\chib + d\t
(\gamma)d\vb  +d\chi (\delta)d\vb \nonumber\\
& &\mbox{}+  d\t   (y) d\tb   +d\t ( r)d\chib +d\chi (s) d\tb +  d\chi
(t)d\chib.
\end{eqnarray}
More precisely,
\beq
||\omega_{A\bar B}||=
\left(
\begin{array}{@{}ccc@{}}  a&  \alpha&  \beta\\   \gamma&  y&  r\\
 \delta&  s &  t
\end{array}
\right); \label{omeg}
\eeq
the entries of $||\omega_{A\bar B}||$ are as follows:
\begin{eqnarray}
 a&=&-{i\over x^2}\left[2\tau+i (\tau-b)  {1+|z|^2\over x}\,
\tb\t +  i(\tau+b)   {1+|z|^2\over x}\,  \chib\chi -
 {\tau^2-b^2\over \tau}  {1+2|z|^2\over x^2}\,
\tb\chib\chi\t \right]; \nonumber \\
\alpha&=& -{1\over x^2}(\tau-b)\vb\t\left[1+ i {\tau+b\over \tau}
 {\chib\chi\over x}  \right]; \qquad \gamma= -{i\over x^2} (\tau-b)z\tb
\left[1+ i{\tau+b\over \tau}
 {\chib\chi\over x}  \right]; \nonumber  \\
\beta&=& -{1\over x^2}(\tau+b)\vb\chi\left[1+i {\tau-b\over \tau}
{\tb\t\over x}  \right];
\qquad \, \delta= -{i\over x^2}(\tau+b)z\chib \left[1+i {\tau-b\over \tau}
 {\tb\t\over x}  \right];  \label{eqC.3}  \\  y&=& -{i\over x}(\tau-b)
\left[1+ i{\tau+b\over 2\tau}  {\chib\chi\over x}  \right]  ; \hskip1.25cm
t=  -{i\over x}(\tau+b)
\left[1+i {\tau-b\over 2\tau}  {\tb\t\over x} \right]; \nonumber \\  r&=&
-{\tau^2-b^2\over 2\tau x^2}\tb\chi  ;\hskip2cm s=  -{\tau^2-b^2\over
2\tau x^2}\chib\t; \hskip2cm    x=1-|z|^2. \nonumber
\end{eqnarray}
A simple
computation leads then to:
\beq
d\mu (z,\vb, \t,\tb,\chi,\chib)=
{-2\tau\over \pi (\tau^2-b^2)}\ dz\, d\vb\, d\t\, d
\tb\, d\chi\, d\chib\ . \label{medida}
\eeq
Clearly, this measure is only valid in the typical case,
$|b|<\tau$. (For the atypical case see Section~6.)
Moreover, the choice of
normalization  in (\ref{medida}) is not innocent.  Its
usefulness will appear in
Section~5 (see Remark 5.4).  The $\OSp$-invariance
of $d\mu$ is claimed
without proof.  In fact, we can show that this is true
using the action of
$\OSp$ on
$\dud$.  The latter can be derived by integrating the flows of the
Hamiltonian vector fields (\ref{vectors})  (see Section~7, 7.5).

It is worth mentioning at this point that up to a slight
adaptation of
conventions (see Appendix~A), the supermeasure
(\ref{medida}) is exactly
the one used in \cite{Balal2} in order to prove the resolution
of the identity
for the typical $\OSp$-CS.  The same result holds here.  In our
notation, this
means that the typical
$\OSp$-CS (\ref{csb}) form an overcomplete basis of $\vtaub$   (see
Section~7, 7.1).  From a computational point of view the way we
evaluate
here (\ref{medida}) is by far simpler than the one used in \cite{Balal2}.

%%%%%%%%%%%%%%%%%%%%%%%%%%%%%%%%%
%%%% 4. Supergeometry and Rothstein's data %%%%%%%
%%%%%%%%%%%%%%%%%%%%%%%%%%%%%%%%%%
\section{More about supergeometry}\label{sec4}

The theory of supermanifolds was originally introduced in
order to provide
physicists with a rigorous framework for studying
supersymmetric field
theories.  Here we are interested in later developments
of the theory that
were oriented towards super extending such techniques
as symplectic
geometry and geometric quantization.  We start then this
section by
presenting a brief account of the key contributions in that
direction
(Section~4.1). The results of Section~3 will then be
rediscussed in the light
of the general theory (Section~4.2 and 4.3).

It is worth stressing that the most important point of this
section is the
extension of Rothstein's characterization of supersymplectic
supermanifolds
to the K\"ahler context (Section~4.3).  This extension was
already discussed
in \cite{Amine1,Amine2} where a definition and a non-trivial
example of
the notion of a super K\"ahler supermanifold were exhibited.
Here, that
definition is made more precise, explicit general formulae are
given, and
another non-trivial example is discussed, namely
$(\dud,\omega)$.

\subsection{Supersymplectic  supermanifolds}

It is known that the geometry of a manifold can be recovered
from its structure sheaf, i.e. the algebra of functions on that
manifold. A
supermanifold is defined by extending  such an algebra to a
supercommutative superalgebra.  The supergeometry of this
supermanifold
is then extracted from the superstructure sheaf thus obtained
through
known techniques \cite{Kost2}.  As for usual manifolds, three
types of
supermanifolds emerge, namely, the $C^\infty$, the real-analytic
and the
complex-analytic (or holomorphic) supermanifolds.  From the
results of the
previous section it clearly appears that we are dealing here
with the last
type of supermanifolds.

\medskip
\noindent{\bf Definition 4.1:} \cite{Green,Roth2}
{\sl A $(p|q)$-dimensional
{\it holomorphic supermanifold\/}
 is a pair
$(M,\am)$, where $M$ is a
$p$-dimensional complex manifold with holomorphic
structure sheaf $\om$
and
$\am$ is a sheaf of supercommutative superalgebras
on $M$, such that:
\begin{description}
\item[(a)]
$\am/\nil$ is isomorphic to $\om$,
$\nil$ being the subsheaf of nilpotent elements of $\am$, and
\item[(b)]
$\am$ is locally isomorphic to the exterior sheaf $\bigwedge\!\she$,
where
$\she\equiv\nil/\nil^2$ is a locally free sheaf over $\om$;  equivalently,
for $\{U_\alpha\}$ an open cover of $M$, $\am(U_\alpha)$ is locally
isomorphic to
$\om(U_\alpha)\otimes\bigwedge\!\C^q\equiv\bigwedge\!\she(U_\alpha)$.
\end{description}\par}
\noindent Here $\bigwedge\!\C^q$ stands for the exterior algebra on
$\C^q$, and
$p$ (resp. $q$) is the even (resp. odd) complex dimension of $(M,\am)$.
Moreover, local supercoordinates on $(M,\am)$ are given by a set
$(z^1,\ldots, z^p; \theta^1, \ldots, \theta^q)$, where   $(z^1,\ldots, z^p)$
are
local coordinates on $M$ and  $(\theta^1, \ldots, \theta^q)$ form a basis of
$\she$ over $\om$.  Notice that up to obvious modifications,
Definition~4.1. applies equally well to $C^\infty$ and real-analytic
supermanifolds.

It is well known that a holomorphic vector bundle over $M$ is completely
specified given its sheaf of holomorphic sections, which is a locally free
sheaf over
$\om$.  Hence,
$\she$ in Definition~4.1 represents the sheaf of sections
of a rank-$q$
holomorphic vector bundle $\F$ over $M$.  To any holomorphic
supermanifold $(M,\am)$ one can then canonically associate the
holomorphic supermanifold
$(M,\om(\bigwedge\!\F))$, where $\om(\bigwedge\!\F)$ is the sheaf of
sections of the exterior bundle $\bigwedge\!\F\rightarrow M$.  Condition
{\bf(b)} above implies that $(M,\am)$ and
$(M,\om(\bigwedge\!\F))$ are
locally isomorphic. The same holds true for
$C^\infty$ and real-analytic
supermanifolds.  In these two last instances the local
isomorphism  always extends to a global but non-canonical
one \cite{Batche}.  However,  this is not always true in
the holomorphic
case \cite{Green,Roth2}.

As for usual manifolds, the tangent sheaf is defined as the sheaf of
superderivations of $\am$, and the cotangent sheaf
$\Omega^1(\am)$ as its
dual.  Then a super de Rham complex can be constructed by
introducing a
coboundary operator $d$.  From these ingredients one
defines the notion of
a supersymplectic supermanifold.

\medskip
\noindent{\bf Definition 4.2:} \cite{Kost2} {\sl A {\it supersymplectic
supermanifold\/} is a triple $(M,\am,\omega)$, where
$\omega$ is a closed and non-degenerate even $2$-superform
on $(M,\am)$.
\par}
\medskip

Rothstein's characterization of $C^\infty$ supersymplectic
supermanifolds in terms of the geometry of vector bundles over usual
symplectic manifolds constitutes a major contribution to this topic
\cite{Roth1}.  Indeed, for
$\omega$ {\it at most quadratic in the odd coordinates\/}, Rothstein's
theorem
\cite{Roth1} allows one to completely identify $\omega$ in terms of a
symplectic structure on
$M$ and extra structures in the vector bundle sector.  More precisely,
using the global isomorphism mentioned above, it states that to any
$C^\infty$ supersymplectic supermanifold
$(M,\am,\omega)$ there corresponds a set $(M,\omega_0, \E, g, \nabla^g)$,
where
$(M,\omega_0)$ is a symplectic manifold, $\E$ is a vector bundle over $M$
with metric $g$ and
$g$-compatible connection $\nabla^g$, such that $\she$ is the sheaf of
linear functionals on $\E$ and $\omega$ is completely determined in terms
of $(\omega_0,g,\nabla^g)$ as follows:
\beq
\omega=\omega_0-d\alpha_2,\quad \hbox{where} \quad
\alpha_2=-g_{ab}\t^a D\t^b.\label{Ro1}
\eeq
Here $D$ is an operator defined on $\bigwedge\!\she$, with values in
$\Omega^1(\am)$, such that:
\beq  D\theta^a\equiv d\theta^a-A^a_{ib}\theta^b dx^i,
\qquad a, b\in\{1,\ldots,q\}, i\in\{1,\ldots,p=2n\},\label{Ro2}
\eeq
where $(x^i;\t^a)$ are now real supercoordinates on $(M,\am,\omega)$, and
$A^a_{ib}$ are the components of $\nabla^g$ in the basis of the generators
${\t^a}$ of
$\she$.  The explicit form of (\ref{Ro1}) is given by:
\beq
\omega=\omega_0+ \demi g_{ab}R^b_{ijc}\theta^c\theta^a dx^i dx^j
+g_{ab}D\theta^a D\theta^b, \label{Ro3}
\eeq
where $R^b_{ijc}$ are the components of curv$\nabla^g$.

The correspondence mentioned above is one to one only in the
$C^\infty$
case.  In the complex-analytic case either one considers supersymplectic
holomorphic supermanifolds in the form
$(M,\om(\bigwedge\!\F),\omega)$,
for $\F$ a holomorphic vector bundle over
$M$, or uses only the one way
correspondence of Rothstein's theorem
\cite{Roth1}. In both situations,
equations similar to (\ref{Ro1})--(\ref{Ro3})
hold; they are explicitly
derived in  Section~4.3.  In what follows,
{\it Rothstein's data\/} will
refer to the set
$(M,\omega_0,
\E, g, \nabla^g)$ associated to a supersymplectic
supermanifold
$(M,\am,\omega)$.

Few precisions are now in order.  To be able to
decompose without
any ambiguity  an even two-superform as a sum
of homogeneous
components in the anticommuting variables, one
needs to restrict bundle
automorphisms of
$\bigwedge\!\she$ to those automorphisms induced
from bundle
automorphisms of
$\she$ \cite{Roth1}.  Moreover, when $\omega$ contains
terms of
higher order in the odd coordinates (more than quadratic),
 the second
part of Rothstein's theorem \cite{Roth1} states that there
exists a
superdiffeomorphism
$\rho$ of
$(M,\am)$ such that
$\rho$ is the identity modulo $\bigwedge^2\!\she$, and
$\rho^*(\omega)$
is at most quadratic in the odd coordinates.  Hence, the first
part of Rothstein's theorem can be applied to $\rho^*(\omega)$.  In
other words, for
$\omega=\omega_0+\omega_2+\omega_4+\dots$, where the
subscripts
refer to the degree of homogeneity in the odd coordinates,  one
needs first to find the superdiffeomorphism $\rho$ (which
depends only on $\omega_4+\dots$), then one uses it in order to
transform $\omega$ into a $2$-superform at most quadratic in
the odd coordinates.  Then one can identify Rothstein's data for
$\omega$.  On the other hand, given those data one cannot
reconstruct the original $2$-superform. Indeed, only the
transformed one
is at reach, since $\rho$ cannot be deduced from the above data.  The
supersymplectic form considered here (\ref{sform1})--(\ref{sform2}), is
obviously of the preceding form, nevertheless we will show in
Section~4.3 that $\omega_4$ can be obtained
explicitly from Rothstein's data (as $\omega_0+\omega_2$),
without any
mention to $\rho$ (see (4.18)).  In other words, the whole
supersymplectic
form can be obtained from the simple knowledge of
$(M,\omega_0,
\E, g, \nabla^g)$.  This seems to be a
common feature of super coadjoint orbits of the type
considered here.

Our task is now twofold.  First, identify $\dud$ as a
holomorphic supermanifold, then identify Rothstein's data for the
supersymplectic
$(\dud,\omega)$. The first part is straightforward.
For the second one, we
will make use of a very useful Lemma proved in \cite{Amine1}
in the case
of $\duu\equiv{\rm OSp}(1/2)/{\rm U}(1)$, and which applies to more
general situations, in particular to the one in hand.

\subsection{$\dud$ as a holomorphic
supermanifold}

As a holomorphic supermanifold, the $\OSp$-homogeneous
superspace
$\dud$ obtained in Section~3 corresponds to the pair
$(\du, {\cal A}_{\du})$, where $\du$ is the unit disc, and ${\cal
A}_{\du}\equiv{\cal A}^{(1|2)}={\cal O}_{\du}\otimes\bigwedge\!\C^2$ is
the defining superstructure sheaf. A general section $h$ of $\aud$ is a
superholomorphic function, i.e.
\beq h(z,\t,\chi)\equiv h_1(z)+\t\,h_2(z)+\chi\,h_3(z)+\chi\t\,h_4(z),
\eeq where the $h_i$'s are holomorphic functions on $\du$.
Moreover,
the vector bundle
$\F$ is nothing but a trivial rank 2 holomorphic vector bundle over
$\du$.

Finally, notice here that the super observables in (\ref{clobs}) are
sections of the complexified superstructure sheaf $\aud_{\C}\equiv
C^\infty_{\C}\otimes\bigwedge\!\C^4$; and the Hamiltonian vector
superfields in (\ref{vectors}) are derivations of $\aud_{\C}$.  The
canonically associated vector bundle is
$\F\oplus\overline {\F}$, where $\overline {\F}$ is the complex
conjugate
bundle of $\F$.

\subsection{$\dud$ as a super K\"ahler
supermanifold}

Let us now identify the data
$(M,\omega_0,\E,g,\nabla^g)$ for
$(\dud,\omega)$.  First of all, it is straightforward from
(\ref{sform2}),
that the symplectic manifold
$(M,\omega_0)$ is here the symplectic unit disc
 $(\du,\omega_0)$, where
$\omega_0\equiv-2i\tau (\de)^{-2} dz\,\! d\zb$.  Indeed, observe that
$\hbox{SU}(1,1)\times \hbox{U}(1)$ is the body of
$\OSp$, and hence the body of
$\dud\equiv \OSp/(\hbox{U}(1)\times \hbox{U}(1))$ is
$\hbox{SU}(1,1)/\hbox{U}(1)\equiv
\du$.  On the other hand, when one sets the odd variables to zero,
the $\OSp$-CS
become the SU$(1,1)$-CS, the underlying geometry of
which is known to
be given by $(\du,\omega_0)$.   Furthermore, according to
Rothstein's
theorem the vector bundle $\E$ is just $\F^*$, the dual of $\F$.

The identification of the remaining data is highly simplified
if one makes
use of the results of \cite{Amine1}.   Indeed, in \cite{Amine1}
it has been
shown that Rothstein's data of the OSp$(1/2)$ coadjoint orbit
studied
there,  can be directly read off from the superfunction $f$
generating the
supersymplectic superform.  More precisely, if one writes $f$ as
$f_0+f_2+f_4+\ldots$, where $f_{2n}$ designates that component of $f$
which is of order $2n$ in the odd coordinates, then it appears that $f_2$
assumes the following form:
\beq f_2=-ig_{a\bar b}\t^a\tb^b,\label{f2}
\eeq
where $g$ is Rothstein's data metric on $\E$; the $\t^a$'s are the
odd supercoordinates of the holomorphic supermanifold considered,
which
can also be viewed as a (local) frame field of $\F$ over $\du$.  The
notations in (\ref{f2}) are those commonly used in complex
geometry, see for example \cite{Koba}.

At this point it is worth anticipating by mentioning that the above
considerations are valid only in the {\it particular complex-analytic\/}
case of a {\it super K\"ahler\/} supermanifold, a notion defined in
\cite{Amine1} and rediscussed below.  Both $\duu$ \cite{Amine1} and
$\dud$ are non-trivial examples of such a notion.

We now write the explicit form of $g$ for $\dud$.  If the odd coordinates
$\t$ and $\chi$ are now denoted, respectively, by $\tu$ and $\td$, then
$f_2$ in (\ref{pot}) is given by:
\beq f_2=i(\tau-b){\tu\tub\over(\de)}+i(\tau+b){\td\tdb\over(\de)}.
\label{f2p}
\eeq
A simple comparison of (\ref{f2}) and (\ref{f2p}) leads to the
following matrix form $||g||$ of $g$ in the frame field of
$\E$ over $\du$, which is dual to that of
$\F$ given above:
\beq ||g||=
\left(
\begin{array}{@{}ccc@{}}  \displaystyle{\tau-b\over(\de)}&0\\     0&
 \displaystyle{\tau+b\over(\de)}
\end{array}
\right).\label{metric}
\eeq
This is a diagonal metric which is clearly  Hermitian \cite{Koba} since we
are considering $|b|<\tau$ (typical CS).

It remains now to identify the connection $\nabla^g$ compatible with this
Hermitian metric.  In order to exhibit it, we rederive Rothstein's results
(\ref{Ro1})--(\ref{Ro2}) in our {\it particular complex-analytic\/}
setting.  Once again this task is highly simplified, thanks to the
observations of \cite{Amine1} that led to formula (\ref{f2}).  We recall
that Rothstein's formulae (\ref{Ro1})--(\ref{Ro3}) were derived in the
$C^\infty$ case
\cite{Roth1}.

The complex-analytic counterpart of the even $1$-superform
$\alpha_2$ appearing in (\ref{Ro1}), can be obtained from
(\ref{f2}) as follows:
\beq
\alpha_2\equiv-{i\over2}(\delta-\bar\delta) f_2.\label{alpha21}
\eeq
A direct computation based on (\ref{f2}), with $g$ an Hermitian
metric on
a holomorphic vector bundle $\E\rightarrow M$, leads to:
\beq
\alpha_2=\demi g_{a\bar b}(\t^a D\tb^b+\tb^b D\t^a),
\label{alpha22}
\eeq
where now,
\beq
D\t^a=d\t^a+\Gamma^a_{ib}\t^b\,\!dz^i, \qquad
D\tb^a=d\tb^a+\Gamma^{\bar a}_{\bar\imath \bar
b}\tb^b\,\!d\bar z^i,\label{alpha23}
\eeq
and
\beq
\Gamma^a_{ib} =  g^{a\bar c}{\partial g_{b\bar c}\over
\partial z^i}.\label{alpha24}
\eeq
One easily recognizes here the $\Gamma^a_{ib}$'s as the
components of
the (canonical) {\it Hermitian connection\/} associated to
the Hermitian
metric
$g$ on $\E$ \cite{Koba}.  Notice that these connection
components are
expressed in the frame field of $\E$, while in the $C^\infty$
case of
(\ref{Ro1})--(\ref{Ro3}) the
$A^a_{ib}$'s are expressed in the frame field of
$\F$, dual to that of $\E$.  As an endomorphism of a fibre,
one is minus
the transpose of the other \cite{Koba}.  This explains the
sign difference
and the change in the notation.

Before stating the final result concerning $(\dud,\omega)$,
let us carry
over and evaluate the complex-analytic counterpart of
(\ref{Ro3}).   From
(\ref{alpha22})--(\ref{alpha24}), a straightforward
computation gives,
\beq
d\alpha_2= g_{a\bar b}D\t^a
D\tb^b- g_{a\bar b}R^a_{i\bar\jmath c}\t^c\tb^b
dz^id\bar z^j,\label{alpha25}
\eeq
where
\beq
R^a_{\bar\jmath ic}\equiv{\partial\Gamma^a_{ic}\over\partial\bar
z^j}.\label{alpha26}
\eeq

{}From what precedes, we have the following immediate result:
\beq
\omega_2=-d\alpha_2,\label{alpha27}
\eeq
where $\omega_2$ is given in
(\ref{O2}) and $d\alpha_2$ is (\ref{alpha25}) for $g$ the Hermitian
metric in (\ref{metric}).
Hence, in summary, we have proven the following:

\medskip
\noindent{\bf Theorem 4.3:} {\sl Rothstein's data for
$(\dud,\omega)$ are:
$(\du,\omega_0, \E, g, \nabla^g)$ where $\du$ in the unit disc,
$\omega_0$ is the SU$(1,1)$-invariant two-form on $\du$
(\ref{O0}),
$\E=\F^*$ is a rank $2$ trivial holomorphic vector bundle
over $\du$, $g$
is the Hermitian metric (\ref{metric}) on $\E$ and $\nabla^g$
is the
corresponding (canonical) Hermitian connection (the
components of which
can be explicitly evaluated using (\ref{alpha24})).\par}
\medskip

Notice now that the symplectic $(\du,\omega_0)$ is
moreover a K\"ahler
manifold, since
$\omega_0=-i\partial\bar\partial f_0$, where $f_0$, the
odd-coordinates-independent part of $f$ in (\ref{pot}), is
a K\"ahler potential for
$\omega_0$; $\partial=dz{\partial\over\partial z}$.  Hence,
one clearly
sees that
$(\dud,\omega)$ is a non-trivial example of the following
definiton of a
super K\"ahler supermanifold:

\medskip
\noindent{\bf Definition 4.4:} {\sl A super K\"ahler
supermanifold
$(M,
\am,
\omega)$ is a holomorphic supersymplectic
supermanifold, whose
Rothstein's data, $(M,\omega_0, \E, g, \nabla^g)$, are such that
$(M,\omega_0)$ is a K\"ahler manifold,
$(\E,g)$ is a holomorphic Hermitian vector bundle over $M$ and
$\nabla^g$ is the canonical Hermitian connection.}\par
\medskip

As already mentioned the present situation
allows us to go beyond
Rothstein's theorem \cite{Roth1}.  Indeed,
$\omega_4$ in
(\ref{sform1})--(\ref{sform2}) can also be
rewritten in terms of
Rothstein's data of $(\dud,\omega)$ obtained
above.  This is achieved by
simply noticing that in (\ref{pot}):
\beq
f_4={1\over 4\tau}(f_2)^2.\label{obs}
\eeq
Then,
\beq
\alpha_4\equiv-{i\over2}(\delta-\bar\delta)f_4
={1\over2\tau}f_2\alpha_2,
\eeq
such that,
\beq
\omega_4\equiv-d\alpha_4=
{1\over2\tau}\left[f_2\omega_2-df_2\alpha_2\right].
\eeq
A simple computation based on (\ref{f2}), (\ref{alpha22}) and
(\ref{alpha25}), leads to the following:
\beq
\omega_4={i\over2\tau}g_{a\bar b}
g_{c\bar d}\left[\t^a\tb^d D\t^c
D\tb^b+\t^a\tb^b\left(D\t^c
D\tb^d-R^c_{i\bar\jmath e}\t^e\tb^d dz^i d\bar
z^j\right)\right].
\eeq
When $g$ is the Hermitian metric in (\ref{metric}),
$\omega_4$ above is
exactly the $\omega_4$ of (\ref{sform2}).  The present
extension of Rothstein's theorem is mainly based on
the observation made
in (\ref{obs}).  This is in fact an intrinsic property of
super K\"ahler coadjoint orbits of some simple Lie
supergroups.
For such orbits, Rothstein's data determine the
complete super K\"ahler
form.  A detailed description of a general framework
will be given
elsewhere.

In analogy with the non-super case, and in view of
(\ref{sform1}) it is
worth calling the real superfunction $f$ in (\ref{pot})
a super K\"ahler
potential for the super K\"ahler form $\omega$ given in
(\ref{sform1})--(\ref{sform2}).  Such a potential  is
defined up to the
addition of a superholomorphic or/and an
anti-superholomorphic function
on $\dud$.  Clearly, Rothstein's data
for a super K\"ahler supermanifold are encoded
in its super K\"ahler
potential $f=f_0+f_2+\ldots$.  Indeed, the body $f_0$
of $f$ is a K\"ahler
potential for the K\"ahler manifold $(M,\omega_0)$,
while $f_2$ provides
the Hermitian  structure on the holomorphic bundle
$\E$ (see \ref{f2}).

The description of super K\"ahler geometry given
above  is sufficient for
our purpose.  However, it is worth mentioning that
Definition~4.4 can be made much more precise.  Indeed,
a deeper analysis
of Rothstein's data for $(\dud,\omega)$ shows that they
determine an
Einstein-Hermitian vector bundle \cite{Koba}
(see Section~7, 7.3).  This
very interesting observation deserves further investigations.
More
details will be given in a forthcoming publication.

The next section addresses the geometric quantization of
$(\dud,\omega)$.
The super K\"ahler character of the latter leads naturally
to the existence
of a super K\"ahler polarization, which makes the
complete quantization
program successful.

%%%%%%%%%%%%%%%%%%%%%%%%%%%%%%%%%
%%%% 5. Geometric quantization %%%%%%%%%%%%%%%
%%%%%%%%%%%%%%%%%%%%%%%%%%%%%%%%%%
\section{Geometric quantization}\label{sec5}

\noindent By {\it geometric quantization\/} we refer here to the
celebrated method, independently devised by Kostant
\cite{Kost1} and
Souriau \cite{Souriau}, which associates to a given classical mechanics a
quantum counterpart.  Kostant-Souriau quantization procedure meets
Kirillov's {\it method of orbits\/} \cite{Kiri1} when the classical
mechanics is described by a coadjoint orbit of a Lie group $G$.  The
quantum output comes then in the form of a Unitary Irreducible
Representation (UIR) of the considered group $G$.  This is the case we are
interested in super extending.  From now on we will focus on this kind of
situations.

In practice, geometric quantization proceeds in two steps.
The first one,
called {\it prequantization\/}, consists in exhibiting a complex line
bundle over $(M\equiv G/H, \omega)$, with Hermitian structure and
compatible connection
$\nabla$, such that curv$\nabla=\omega$.  Such a line bundle exists
whenever $[\omega]$ is an integral cohomology class
(integrality condition).  When
lifted to this line bundle, the transitive (and symplectic)  action of
$G$ on
$M$ gives rise to a unitary but reducible representation of $G$.
The second
step consists in using a {\it polarization\/} in order to select an
irreducible subrepresentation of the prequantum representation.  More
precisely, the group action is restricted to the subspace of those $L^2$
sections of the prequantum line bundle which are covariantly constant
along the vector fields generating the polarization.  More details
concerning the general procedure of geometric quantization can
be found in
\cite{Wood,Kiri2}.

Super-prequantization was partly developed by Kostant \cite{Kost2}.
Tuynman \cite{Gijs1} completed that construction by equipping Kostant's
super-prequantum bundle with a super Hermitian structure compatible
with the connection.  We will follow here that construction, assuming
that the reader is at least familiar with Kostant's work.  On the other
hand, because of the lack of a notion of polarization, the second part of
the program was not considered in
\cite{Kost2}.  Here, as a super K\"ahler supermanifold $(\dud,\omega)$ is
naturally equipped with a {\it super K\"ahler polarization\/} that allows
us to carry out the whole quantization program.

Before giving the details of the construction, it is worth mentioning that
the notion of a polarization in the super context and in connection
with geometric quantization already appeared in the literature.  For
instance, in the real case, a general definition of a polarization was given
in \cite{Gijs1};  it was then used to quantize the BRST charge.  On the
other hand, a super K\"ahler polarization was introduced in \cite{Schaller}
in order to quantize a field theory with fermionic
degrees of freedom, i.e. an infinite dimensional flat phase space.
Here we use a general notion in the complex-analytic case which allows
quantizing non-trivial super phase spaces, such as coadjoint orbits.

Finally, let us mention that recently another quantization method, namely
the deformation quantization, has been extended to the super context in
\cite{BKLR}; this method was applied in particular to the super unit disc
$\duu$.  Geometric quantization of
the latter is considered in \cite{Amine2}.

\subsection{Super prequantization}

Following Kostant's general scheme \cite{Kost2}, in order to
prequantize $(\dud,\omega)$, one needs to exhibit a complex {\it line
bundle sheaf\/}
$\lud$ with connection $\nabla$ over $\dud$, such that $\hbox{curv}
\nabla=\omega$.  Such a line bundle sheaf exists if and only if there
exists a
complex line bundle $\L$ with connection $\nabla_0$ over $\du$, such that
$\hbox{curv}
\nabla_0=\omega_0$ (i.e. {\it iff\/} $(\du,\omega_0)$ is prequantizable)
\cite{Kost2}.  Since our $\omega_0$ is exact, it is known that such an $\L$
always exists.  It is then not hard to see that,
\beq
\lud\equiv\lu\otimes{\cal A}^{(1|2)}_{\C} \cong{\cal
A}^{(1|2)}_{\C},
\label{eq5.1}
\eeq
where $\lu$ is the sheaf of $C^\infty$ sections of $\L$.

To the classical observables in (\ref{clobs}) one can associate the
so-called prequantum operators, which act in the space of $C^\infty$
sections of $\lud$. These operators are obtained using the
following formula \cite{Kost2},
\beq
\hat H\equiv-i\nabla_{X_H}+H,\quad
\hbox{where}\quad\nabla_{X_H}=X_H-iX_H\pint\vartheta,   \label{eq5.2}
\eeq
and
\begin{eqnarray}
\vartheta& \equiv & -i\delta f = - 2i\tau{\zb dz \over 1-\mz}
\left[ 1+i  {\tau -b \over 2\tau }    {\tb\t\over 1-\mz}   +i   {\tau +b \over
2\tau }    {\chib\chi \over 1-\mz}  - {\tau^2 -b^2 \over 2\tau^2 }
{\tb\chib\chi\t \over (1-\mz)^2} \right] \nonumber \\ & & +(\tau-b) {d\t\
\tb  \over 1-\mz} \left[1+i {\tau +b \over 2\tau }  {\chib\chi
\over 1-\mz}  \right]  +(\tau+b){ d\chi\ \chib  \over 1-\mz}  \left[1+i
{\tau -b \over 2\tau }    {\tb\t\over 1-\mz}  \right]
\label{eq5.3}
\end{eqnarray}
is a $1$-superform potential for the connection $\nabla$
equipping the prequantum bundle $\lud$; it is such that
$\omega=-d\vartheta$.  A straightforward computation gives,
\begin{eqnarray}
\hat B &= & \displaystyle {\t\over 2 }\pt -
{\tb\over 2}\ptb -  {\chi\over2}
\pchi +  {\chib\over 2} \pchib +b \ ,  \nonumber   \\
\hat K_0 &=& \displaystyle  z\pv-\vb\pvb+
{\t\over 2} \pt -  {\tb\over 2} \ptb +
{\chi\over 2} \pchi -  {\chib\over 2} \pchib + \tau  \ ,
 \nonumber \\
\hat K_+ &=&\displaystyle  z^2\pv-\pvb+z\t\pt+z\chi\pchi  +
2\tau z\ ,\nonumber\\
\hat K_- &=& \displaystyle \pv-\vb^2\pvb-\vb\tb\ptb-
\vb\chib\pchib  \ ,
\label{hat}\\
\hat V_+ &=& \displaystyle -i\left( {\tau-b\over 2\tau}
\right) z\t\pv -  \left(
{\tau+b\over 2\tau}\right) \chib\pvb +\ptb +i \left(z+
\left(  {\tau-b\over
2\tau}\right) \chi\t\right) \pchi  -i(\tau-b)\t   \ ,
\nonumber   \\
\hat W_- &=&  \displaystyle -i
\left( {\tau+b\over 2\tau}\right) \chi\pv -
\left(
{\tau-b\over 2\tau}\right) \vb\tb\pvb +
i\pt +\left(\vb- \left(  {\tau-b\over
2\tau}\right)\tb \chib\right) \pchib  \ ,
\nonumber  \\
\hat V_- &=& \displaystyle -i\left(
{\tau-b\over 2\tau}\right) \t\pv - \left(
{\tau+b\over 2\tau}\right) \vb\chib\pvb  +
\left(\vb + \left({\tau+b\over
2\tau}\right) \tb\chib\right) \ptb +i\pchi   \ ,
\nonumber   \\
\hat W_+ &=& \displaystyle -
i\left( {\tau+b\over 2\tau}\right) z\chi\pv -
\left( {\tau-b\over 2\tau}\right) \tb\pvb  +
i\left(z-
\left({\tau+b\over 2\tau}\right) \chi\t\right) \pt+\pchib
-i(\tau+b)\chi \ .
\nonumber
\end{eqnarray}
One easily verifies that these operators close to the $\osp$
Lie superalgebra.  They provide thus a representation of $\osp$ in the
space of $C^\infty$ sections of $\lud$.

In the next subsection, using a natural
{\it invariant super K\"ahler polarization\/} on $\dud$ we will select a
subsheaf
$\ludp$ of $\lud$; $\ludp$ will then  be equipped with a
$\nabla$-compatible {\it
super Hermitian structure\/}.  When restricted to $\ludp$
the above prequantum
representation will reduce to a {\it super unitary irreducible
representation\/}
of $\osp$.

\subsection{Super polarization}

As for the K\"ahler unit disc $\du$, a natural super polarization,
called here {\it super K\"ahler\/} and denoted ${\cal P}$, exists on $\dud$.
It is spanned by the vector superfields $\pz, \pt$ and
$\pchi$.   One easily verifies that this is actually a good candidate for a
super polarization.  Indeed, ${\cal P}$ fulfills all the required conditions,
namely,
\begin{description}
\item[(i)] ${\cal P}$ is involutive;
\item[(ii)] ${\cal P}$  is maximal isotropic, i.e. $\omega(Z,Y)=0,
\forall\ Z, Y \in {\cal P}$ and dim\,${\cal P}=(1|2)=\,$dim\,$\dud$ as a
holomorphic supermanifold.
\end{description} Moreover,  one can easily verify that,
\begin{description}
\item[(iii)] ${\cal P}\cap\overline{\cal P}=\{0\}$;
\item[(iv)] ${\cal P}$ is invariant, namely, for $X_H$ one of the
Hamiltonian vector fields in (\ref{vectors}),
$[X_H,Z]\in{\cal P},\ \forall\ Z\in {\cal P}$ (brackets denote here
a commutator or an anticommutator).
\end{description}
Property {\bf(iii)} confirms, in agreement with the
results of Section~4, that ${\cal P}$ deserves to be called a {\it super
K\"ahler\/} polarization for $\dud$.  On the other hand, property {\bf(iv)}
means that the Hamiltonian flows of the classical observables (\ref{clobs})
preserve ${\cal P}$; the importance of this property will be stressed
soon.   Moreover, as it will be shown at the end of this section,
${\cal P}$ is {\it positive\/}.  This property ensures that the final
representation space is non-trivial. In summary,
${\cal P}$ is a positive invariant super K\"ahler polarization.

This polarization is now used in order to select a subsheaf $\ludp$ of
$\lud$.  The latter consists of the sections
$\psi(z,\zb,\t,\tb,\chi,\chib)$ of
$\lud(\du)$ which are covariantly constant along $\overline{{\cal P}}$, i.e.
those sections $\psi$ which are such that:
\beq
\displaystyle\nabla_{\!\!\pzb}\psi=0\ ,\ \
\displaystyle\nabla_{\!\!{\partial\over\partial
\vphantom{\bar{\bar\theta}}\tb}}\psi=0,
\quad\hbox{and}\quad
\displaystyle\nabla_{\!\!{\partial\over\partial
\vphantom{\bar{\bar\theta}}\chib}}\psi=0,  \label{eq5.5}
\eeq   where, from now on, the covariant superderivation is taken as in
(\ref{eq5.2}) but with $\vartheta$, the super\-polarization-adapted
$1$-superform  potential for
$\nabla$, replaced by the {\it real\/} $1$-superform potential $\alpha$
given by,
\begin{eqnarray}
\alpha&\!\!\!\equiv\!\!\!&-{i\over2}(\delta-\bar\delta)f\nonumber\\
&\!\!\! =\!\!\! &i\tau\, { z\,d\zb -\zb\, dz
\over\de}\left[  1+i  {\tau -b
\over 2\tau }  \,  {\tb\t\over 1-\mz}   +i  {\tau +b \over 2\tau } \,
{\chib\chi
\over 1-\mz}  - {\tau^2 -b^2
\over 2\tau^2 }\, {\tb\chib\chi\t \over (1-\mz)^2} \right]  \label{eq5.6} \\
&\phantom{\!\!=\!\!} & + {\tau-b\over 2}\,  {d\t\, \tb +d\tb\, \t  \over
1-\mz} \left[1+i {\tau +b
\over 2\tau } \, {\chib\chi
\over 1-\mz}  \right]  +{\tau+b\over 2}\,{ d\chi\, \chib + d\chib\, \chi
\over 1-\mz} \left[1+i {\tau -b \over 2\tau }  \,  {\tb\t\over 1-\mz}
\right].
\nonumber
\end{eqnarray}
In (\ref{eq5.6}) $f=f(z,\zb,\t,\tb,\chi,\chib)$ is the super K\"ahler
potential (\ref{pot}).  Solutions to (\ref{eq5.5}) are of the form
\beq
\psi(z,\zb,\t,\tb,\chi,\chib) =  \exp(-f/2)\,
\phi(z,\t,\chi),\label{sholo}
\eeq where
$\phi$ is a superholomorphic section of $\lud$, and
$\exp(-f/2)\equiv{\cal N}'$  (see (\ref{N'}) and (\ref{pot})).  Notice from
(\ref{sholo}) that
$\ludp$ is isomorphic to ${\cal A}^{(1|2)}$, the superstructure sheaf of
$\dud$ viewed as a holomorphic supermanifold.  Moreover,  property {\bf
(iv)} above ensures that the action of the prequantum operators
(\ref{hat}) in $\ludp$ leaves the latter invariant.

Let us now equip $\ludp(\du)$ with a super Hermitian structure
$(\,\cdot\,,\,\cdot\,)$.  For $\psi=\psi_{\bar 0}+\psi_{\bar1}$ and
$\psi'=\psi_{\bar 0}'+\psi_{\bar 1}'
\in
\ludp(\du)=\ludp(\du)_{\bar 0}\oplus\ludp(\du)_{\bar 1}$, this is given by:
\beq
(\psi',\psi)\equiv\overline{\psi'}\,\psi=\overline{\psi'_{\bar 0}}\,\psi_{\bar
0}+
\overline{\psi'_{\bar 1}}\,\psi_{\bar 1}+\overline{\psi'_{\bar 0}}\,\psi_{\bar
1}+\overline{\psi'_{\bar 1}}\,\psi_{\bar 0}.
\label{SH4}
\eeq
The latter clearly satisfies (\ref{SH2}).  Notice however that
$(\,\cdot\,,\,\cdot\,)$ is {\it not\/} homogeneous, i.e. it is not of the form
(\ref{SH2'}).  Moreover, it is not hard to verify that the super connection
$\nabla$ on
$\lud$ is compatible  with (\ref{SH4}), i.e.
\beq X(\psi',\psi)=(\nabla_X\psi',\psi)+
(-1)^{\epsilon(\psi')\epsilon(X)}(\psi',\nabla_X\psi),\label{comp}
\eeq
where $\psi'$ is now a homogeneous section of $\ludp(\du)$ of parity
$\epsilon(\psi')$, $X$ is a real homogeneous vector superfield on $\dud$ of
parity
$\epsilon(X)$, and $\nabla_X=X-iX\pint\alpha$, for $\alpha$  given in
(\ref{eq5.6}).

As for usual geometric quantization in presence of a K\"ahler polarization,
using the super Hermitian structure (\ref{SH4}), the superspace of
sections
$\psi$ of
$\ludp(\du)$ can be equipped with a super-inner product, given by:
\beq
\langle\!\langle\psi',\psi\rangle\!\rangle_{\tau,b}\equiv \int_{\dud}
(\psi',\psi)\ d\mu,\label{rmk53}
\eeq where $d\mu=d\mu(z,\zb,\t,\tb,\chi,\chib)$ is the $\OSp$-invariant
super measure on
$\dud$ obtained from $\omega$ in (\ref{medida}).  Because of the
isomorphism mentioned after (\ref{sholo}), this super-inner product can
be understood as an inner product on the space of sections $\phi$ of
${\cal A}^{(1|2)}$ (i.e. the space of superholomorphic sections of $\lud$).
Hence, using (\ref{sholo}), we can write
\beq
\langle\!\langle\phi',\phi\rangle\!\rangle_{\tau,b}=
\int_{\dud}  e^{-f}\ (\phi',\phi)\ d\mu.  \label{eq5.9}
\eeq   Let us now investigate the status of this super-inner product from
the Hilbert  space point of view.

Since $|b|<\tau$, one can always write a superholomorphic function
$\phi(z,\t,\chi)$ on $\dud$ as follows:
\beq
\phi(z,\t,\chi)=\phi_1(z)+\t\,\sqrt{\tau-b}\,\phi_2(z)+
\chi\,\sqrt{\tau+b}\,\phi_3(z) +
\chi\t\,\sqrt{{(\tau^2-b^2)(2\tau+1)\over 2\tau}}\,\phi_4(z) ,
\label{eq5.8}
\eeq  where $\phi_i, i=1,\ldots, 4,$ are holomorphic functions on
$\du$.

The super integration in (\ref{eq5.9}) can be partially carried out.  Indeed,
replacing  (\ref{pot}), (\ref{medida}), (\ref{SH4}) and (\ref{eq5.8}) in
(\ref{eq5.9}), and using Berezin integration over the odd Grassmann
variables
\cite{Berezin3}, namely, the only non-zero integral being $\int d\t\,d\tb\,
d\chi\,d\chib\,
\,(\chib\,\chi\,\tb\,\t)=1$, one obtains:
\beq
\langle\!\langle\phi',\phi\rangle\!\rangle_{\tau,b}=
\langle\phi_1',\phi_1\rangle_{k=\tau}+i\,
\langle\phi_2',\phi_2\rangle_{k=\tau+\demi} +
i\,\langle\phi_3',\phi_3\rangle_{k=\tau+\demi}
+\langle\phi_4',\phi_4\rangle_{k=\tau+1}\ \in \C,\label{eq5.10}
\eeq where
\beq \langle\phi',\phi\rangle_k\equiv{2k-1\over\pi}\int_{\du}\
\overline{\phi'(z)}\ \phi(z)\ {dz\,d\zb\over(1-\mz)^{2-2k}}, \quad
\mbox{for}\quad k>\demi, \label{eq5.11}
\eeq  is the usual inner product on the representation space of the
holomorphic (positive) discrete series
$D(k)$ of $\su$, which arise through geometric quantization of the unit  disc
$(\du,\omega_0)$
\cite{dbe,Gijs2,Renouard}.

We can now define a natural notion of square integrability of
superholomorphic sections of a prequantum bundle sheaf, and thus that of
a {\it super Hilbert space\/}.  First observe that
$\langle\!\langle\,\cdot\,,\,\cdot\,\rangle\!\rangle_{\tau,b}$  is an {\it
even super Hermitian form\/} on the space of superholomorphic sections
of $\lud$.  Indeed,
\beq
\overline{\langle\!\langle\phi',\phi\rangle\!\rangle}_{\tau,b}=
(-1)^{\epsilon(\phi')\epsilon(\phi)}\langle\!\langle\phi,
\phi'\rangle\!\rangle_{\tau,b}\ ,
\label{sesqui1}
\eeq  and moreover,
\beq
\langle\!\langle\phi',\phi\rangle\!\rangle_{\tau,b}=
\langle\!\langle\phi'_{\bar 0},\phi_{\bar 0}\rangle\!\rangle_{\tau,b,{\bar
0}} +i \langle\!\langle\phi'_{\bar 1},\phi_{\bar
1}\rangle\!\rangle_{\tau,b,{\bar 1}}\ .\label{sesqui2}
\eeq The new quantities and notation in the above equation can be easily
identified in terms of the old ones appearing in (\ref{sholo}), (\ref{eq5.10})
and (\ref{eq5.11}).  A simple comparison of (\ref{sesqui1}) and
(\ref{sesqui2}) with respectively (\ref{SH2}) and (\ref{SH2'}) confirms our
claim.   This suggests the following natural definition:

\medskip
\noindent{\bf Definition 5.1:} {\sl A superholomorphic section
$\phi=\phi_{\bar 0}+\phi_{\bar 1}$ of
$\lud$ is said to be square integrable if both $\phi_{\bar 0}$ and
$\phi_{\bar 1}$ are respectively square integrable with respect to the
Hermitian forms
$\langle\!\langle\,\cdot\,,\,\cdot\,\rangle\!\rangle_{\tau,b,{\bar 0}}$ and
$\langle\!\langle\,\cdot\,,\,\cdot\,\rangle\!\rangle_{\tau,b,{\bar
1}}$.}\par
\medskip
\noindent  A  definition of a  {\it super Hilbert space\/}, different from
those proposed in \cite{Dewitt,Schmitt,Nagamachi}, immediately follows.

\medskip
\noindent{\bf Definition 5.2:} {\sl A super Hilbert space is a pair
$(\h,
\langle\!\langle\,\cdot\,,\,\cdot\,\rangle\!\rangle)$, where
$\h\equiv\h_{\bar 0}\oplus\h_{\bar 1}$ is a superspace equipped with a
super Hermitian form,
$\langle\!\langle\,\cdot\,,\,\cdot\,\rangle\!\rangle=\langle\!\langle
\,\cdot\,,\,\cdot\,\rangle
\!\rangle_{\bar
0}+i\langle\!\langle\,\cdot\,,\,\cdot\,\rangle\!\rangle_{\bar 1}$, such
that
$(\h_{\bar 0},\langle\!\langle\,\cdot\,,\,\cdot\,\rangle\! \rangle_{\bar
0})$ and
$(\h_{\bar 1},\langle\!\langle\,\cdot\,,\,\cdot\,\rangle\! \rangle_{\bar
1})$ are both Hilbert spaces.}\par
\medskip

According to these definitions, the space of superholomorphic
sections of
$\lud$ equipped with the super Hermitian form  (\ref{eq5.10}) can be
turned into a super Hilbert space simply by taking
$\phi_i, i=1,\ldots, 4$,  to be $L^2$ functions on the unit disc with
respect
to the corresponding Hermitian forms in (\ref{eq5.8}), namely, that
$||\phi_1||^2_{k=\tau}<
\infty$,
$||\phi_2||^2_{k=\tau+\demi}<\infty$,
$||\phi_3||^2_{k=\tau+\demi}<
\infty$ and $||\phi_4||^2_{k=\tau+1}<\infty$.
Hence, in these conditions,
$\phi(z,\t,\chi)$ in (\ref{eq5.8}) can be called an  $L^2$
superholomorphic
section of $\lud$ with respect to the super Hermitian form of
(\ref{eq5.9}).  The obtained super Hilbert space, denoted
$\h_{\tau,b}$,
constitutes then the representation superspace carrying an
{\it explicit\/}
realization of the typical irreducible representation $\vtaub$ of $\osp$.

\begin{description}
\item[Remark 5.3:] We could have equipped $\lud(\du)$ with an  {\it
even\/} super Hermitian structure $(\!(\,\cdot\,,\,\cdot\,)\!)$ instead of
the one of indefinite parity introduced in (\ref{SH4}). The former would
assume the following form
$(\!(\psi',\psi)\!)\equiv\overline{\psi'_{\bar
0}}\,\psi_{\bar 0}+
\overline{\psi'_{\bar 1}}\,\psi_{\bar 1}$.  In this case (\ref{comp}) will be
true only modulo {\it odd\/} quantities.  However,
$\langle\!\langle\,\cdot\,,\,\cdot\,\rangle\!
\rangle_{\tau,b}\equiv\int_{\dud}(\,\cdot\,,\,\cdot\,)\,d\mu=
\int_{\dud}(\!(\,\cdot\,,\,\cdot\,)\!)\,d\mu$, since
$(\!(\,\cdot\,,\,\cdot\,)\!)$ and $(\,\cdot\,,\,\cdot\,)$ differ only by {\it
odd\/} quantities which disappear when one integrates over
anti-commuting variables using Berezin integration.  For the same reason,
all the results following (\ref{rmk53}) will still hold true.
\item[Remark 5.4:] The important result in (\ref{eq5.10})--(\ref{eq5.11})
follows from the special form adopted for $\phi$ in
(\ref{eq5.8}), and the choice of normalization made for $d\mu$ in
Section 3.3, namely, $\int_{\dud}\,\exp(-f)\,d\mu=1$.
\end{description}

The generators of $\osp$ are represented in $\h_{\tau,b}$ by the
super holomorphic restrictions of the prequantum operators (\ref{hat}).
More precisely,
\begin{eqnarray}
\widehat B &= & \displaystyle {\t\over 2 }\pt -  {\chi\over2}
\pchi  +b \ ,  \nonumber   \\
\widehat K_0 &=& \displaystyle  z\pv+ {\t\over 2} \pt +  {\chi\over 2}
\pchi  + \tau  \ ,  \nonumber
\\
\widehat K_+ &=&\displaystyle  z^2\pv+z\t\pt+z\chi\pchi  +2\tau z\
,\nonumber\\
\widehat K_- &=& \displaystyle \pv  \ ,
\label{widehat}\\
\widehat V_+ &=& \displaystyle -i\left( {\tau-b\over 2\tau}\right) z\t\pv
 +i \left(z+ \left(  {\tau-b\over 2\tau}\right) \chi\t\right) \pchi
-i(\tau-b)\t   \ , \nonumber
\\
\widehat W_- &=&  \displaystyle -i\left( {\tau+b\over 2\tau}\right) \chi\pv
+i\pt  \ ,  \nonumber  \\
\widehat V_- &=& \displaystyle -i\left( {\tau-b\over 2\tau}\right) \t\pv
 +i\pchi   \ ,
\nonumber   \\
\widehat W_+ &=& \displaystyle -i\left( {\tau+b\over 2\tau}\right)
z\chi\pv  +i\left(z-
\left({\tau+b\over 2\tau}\right) \chi\t\right) \pt-i(\tau+b)\chi \ .
\nonumber
\end{eqnarray}

The obtained representation is {\it super unitary\/}.  Indeed,  in
agreement with (\ref{comp}), the {\it superadjoint\/} (or {\it super
Hermitian conjugate\/})
$\widehat X^{\dagger}$ of an operator
$\widehat X$ acting in $\h_{\tau,b}$ is defined as follows:
\beq
\langle\!\langle{\widehat
X}^{\dagger}\,\phi',\phi\rangle\!\rangle_{\tau,b}=(-1)^{\epsilon(\phi')
\epsilon(X)}
\langle\!\langle
\phi',{\widehat X}\,\phi\rangle\!\rangle_{\tau,b},\label{sadj}
\eeq for $\phi'$ a homogeneous superholomorphic section of $\lud$ and
${\widehat X}$ a homogeneous operator.  Hence, the quantum counterpart
$\widehat H$ of a {\it real\/} classical observable
$H^{cl}$ is a {\it self superadjoint\/} operator acting in $\h_{\tau,b}$, i.e.
${\widehat H}^{\dagger}=\widehat H$.  Specifically, since both
$B^{cl}$ and
$K_0^{cl}$ are real (see (\ref{reality})), the associated quantum operators
in
(\ref{widehat}) are self superadjoint:
\beq {\widehat B}^{\dagger}=\widehat B\quad \hbox{and}\quad {\widehat
K_0}^{\dagger}=\widehat K_0.\label{sadj1}
\eeq On the other hand, the reality of
$(K^{cl}_++K^{cl}_-)$, $(V^{cl}_++iW^{cl}_-)$ and of similar
combinations that
can be obtained from (\ref{reality}), leads to the following,
\beq ({\widehat K}_+)^{\dagger}={\widehat K}_-, \quad ({\widehat
V}_+)^{\dagger}=i{\widehat W}_-, \quad\hbox{and}\quad ({\widehat
V}_-)^{\dagger}=i{\widehat W}_+.\label{sadj2}
\eeq As it should be, these results are in perfect agreement with the
relations we started with at the level of the abstract representation
theory (compare (\ref{sadj})--(\ref{sadj2}), with (\ref{superA}),
(\ref{superA1}) and (\ref{superA2})).

Geometric quantization of $(\dud,\omega)$ is hence completed .  It
remains now to integrate the obtained super unitary irreducible
representation of
$\osp$ to a representation of $\OSp$ (see Section~7, 7.5).
The  integrability condition, which we assumed from the begining, ensures
that such a procedure leads actually to a non-trivial representation of
$\OSp$.  Indeed, recall that we started with an {\it integrable\/} typical
$\osp$-module
$\vtaub$, i.e.  both $b$ and $\tau$ $\in\demi \N$.  Moreover, because of
(\ref{eq5.11}), we consider  $\tau>\demi$.  This last condition is a
direct consequence of the so-called metaplectic correction to geometric
quantization \cite{Wood}, which we do not consider here.

Finally we briefly discuss the positivity of our super K\"ahler polarization
$\cal P$.  In the non-super context, the positivity of a K\"ahler polarization
ensures that the unitary irreducible representation obtained is non-trivial,
i.e. the corresponding Hilbert space does not reduce to the zero function
\cite{Wood}.  If a super extension of the positivity exists, then ${\cal P}$ is
positive,  since $\h_{\tau,b}$ is clearly non-trivial.  An analysis of the
present
situation and of the one considered in \cite{Amine2} lead to a natural
supergeometric definition of the positivity of a super K\"ahler
polarization.  This will be given elsewhere.

%%%%%%%%%%%%%%%%%%%%%%%%%%%%%%%%%%%%%%
%%%% 6. Atypical OSp(2/2)-CS %%%%%%%%%%%%%%%%
%%%%%%%%%%%%%%%%%%%%%%%%%%%%%%%%%%%%%%
\section{Atypical OSp(2/2)-CS and  associated  orbits}\label{sec6}

We now briefly discuss the main results concerning the
$\OSp$ atypical representations: the atypical coherent states and their
underlying supergeometry.  Recall that the atypical representations occur
when $|b|=\tau$.  It is then not hard to see that by taking
$b\rightarrow
\tau$ or $b\rightarrow -\tau$, almost all the results obtained in the previous
sections reduce to atypical analogs.  Here we emphasize the main
features of the case $b=-\tau$ (one obtains exactly the same results when
$b=\tau$).

If $b=-\tau$, the lowest weight vector is
$|-\tau,\tau,\tau\rangle\equiv|\tau,\tau\rangle$.  As it is shown in
Appendix~B, $W_+|\tau,\tau\rangle$ is a primitive
vector generating an $\osp$-submodule of zero-norm states.  Hence,
$|\tau,\tau\rangle$ satisfies not only (2.5)--(2.6), but it is also such that
\beq W_+|\tau,\tau\rangle=0.\label{W}
\eeq
The appropriate irreducible $\osp$-module is no
longer
$V(\tau,b)$, given in (\ref{vtaub2}), but the one given in (\ref{vtaub4}).
Instead of having four families of states, as in (\ref{vtaub3}) or (\ref{B.4}),
we have only two:
\beq K_+^m |\tau,\tau\rangle \qquad \hbox{and}\qquad  K_+^mV_+^m
|\tau,\tau\rangle ,\ \ m\ge 0.
\eeq In other words,
\beq  U(\tau)\equiv V(\tau,-\tau)/V'(\tau,
-\tau)\equiv\hbox{span}\{|\tau,\tau+m\rangle, |\tau+\textstyle{1\over
2},\tau+\textstyle{1\over 2}+m\rangle, m\in\N\}.\label{vtaub5}
\eeq

The same techniques as in Section~2 lead to the following atypical CS in
$\widetilde{U}(\tau)$. Starting with
$|\tau,\tau\rangle$, these CS appear to be parametrized only by two
variables.  Indeed, as a consequence of (\ref{W}),
\begin{eqnarray}  |a,\t,\chi\rangle&\!\!\!\equiv\!\!\!&{\cal M}\exp(aK_+ +\t
V_++\chi W_+)|\tau,\tau\rangle,\nonumber\\ &\!\!\!\equiv\!\!\!&{\cal
M}\exp(zK_+ +\t V_+)|\tau,\tau\rangle\equiv|z,\t\rangle,\label{ecr}
\end{eqnarray}  where $a\in{\cal B}_0$, and $\t, \chi\in{\cal B}_1$ such that
$z=a-{1\over 2\tau}\chi\t\in \C$.   A simple comparison of (\ref{ecr}) with
the corresponding equation for the typical CS in Section~2, shows that the
atypical CS are simply obtained from the typical ones by taking $b=-\tau$;
clearly they do not depend on $\chi$.  More precisely,
\begin{eqnarray}  |z, \t\rangle & = & {\cal M} \left[\, \sum_{m=0}^{\infty}
\sqrt{\textstyle{\Gamma(2\tau+m)\over m!\, \Gamma(2\tau)}}\ z^m
|\tau,\tau+m\rangle   \right. \nonumber  \\ &  &\qquad\quad \left. + \t
\sqrt{2\tau} \sum_{m=0}^{\infty}
\sqrt{\textstyle{\Gamma(2\tau+m+1)\over m!\, \Gamma(2\tau+1)}}\ z^m
|\tau+\textstyle{1\over2},\tau+\textstyle{1\over2}+m \rangle   \right],
\label{acs}
\end{eqnarray}    where the real normalization constant is
\beq  {\cal M}=(\de)^{\tau}\left[ 1+i\tau\,{\tb\t\over \de} \right]. \label{M}
\eeq

At this point it is worth noticing, in connection with Remark 2.7, that the
atypical CS are nothing but OSp$(1/2)$-CS \cite{Balal1,Amine1},  where
here OSp$(1/2)$ is the subsupergroup of $\OSp$ generated by
$\{K_0, K_\pm, F_\pm\equiv (V_\pm+W_\pm)/{\sqrt2}\}$.  More precisely,
\beq  |z,\t\rangle= {\cal M}\exp(zK_+ +{\sqrt2}\t F_+) |\tau,\tau\rangle;
\eeq   once again this is a direct consequence of (\ref{W}).  Hence, up to the
above rescaling  of $\t$, the analysis of the supergeometry underlying the
$\OSp$ atypical CS, and its geometric quantization  reduce {\it almost\/} to
those already studied in \cite{Amine1,Amine2}.  The main  differences are
discussed below.

The variables $z$ and $\t$ above parametrize the $N=1$ super unit disc
${\rm OSp}(1/2)/{\rm U}(1)\equiv\duu$ \cite{Amine1}.  The super K\"ahler
potential on ${\cal D}^{(1\vert 1)}$ is given by,
\beq  f(z,\vb,\t,\tb) = \log |\langle \tau,\tau|\vb,\tb\rangle |^{-2} =-
2\tau\left[ \log (\de)- i {\tb\t \over \de}\right] ,
\eeq  and the super K\"ahler superform is simply,
\beq
\omega={-2i \tau\over (\de)^2}\left[1+i\tb\t{1+|z|^2\over(\de)}\right]\,  dz
\, d\vb-{2\tau\over
\de}d\t\,  d\tb - {2\tau\over (\de)^2} \left[\t\vb\, dz \, d\tb-\tb z\, d\t \,
d\vb
\right]\!\!.\label{s-form2}
\eeq

The $\osp$ classical observables are obtained from (\ref{clobs}) by taking
$b=-\tau$, or equivalently by formally setting $\chi=0$.  When evaluated as
in the typical case, the Hamiltonian vector superfields
assume now the following form,
\begin{eqnarray}
X_{B{\phantom1}}&\!\!\!=\!\!\!&\displaystyle i \, {\t\over
2 }\,
\pt - i\,  {\tb\over 2} \, \ptb ,\qquad\qquad\enskip\,  X_{K_0} =
\displaystyle  iz\,
\pv-i\vb\, \pvb+i \, {\t\over 2}\,  \pt - i \,  {\tb\over 2} \,  \ptb ,\\
X_{K_+} &\!\!\!=\!\!\!& \displaystyle  iz^2\, \pv-i\, \pvb+iz\t \, \pt,
\quad\    X_{K_-} = \displaystyle i \, \pv-i\vb^2 \, \pvb-i\vb\tb\,
\ptb,\\  X_{V_+}&\!\!\!=\!\!\!& \displaystyle  z\t \, \pv  +i\, \ptb ,
\qquad\qquad\quad\ \,   X_{W_-} =
\displaystyle - i \vb\tb\, \pvb -\pt, \\  X_{V_-} &\!\!\!=\!\!\!&
\displaystyle \t\, \pv  +i\vb\, \ptb, \qquad\qquad\quad\ \, X_{W_+} =
\displaystyle  - i
\tb\, \pvb  -z \, \pt .\label{isotropy}
\end{eqnarray}
This corresponds to simultaneously taking $b=-\tau$ and
eliminating
$\chi$ in (\ref{vectors}).

{}From these explicit expressions one immediately identifies the superalgebra
of the isotropy subsupergroup $G_0$ at the origin $(z=0, \t=0)$ of the
atypical phase space.  Clearly, $X_B$, $X_{K_0}$, $X_{V_-}$ and
$X_{W_+}$ act trivially at the origin.  Hence, the isotropy subsupergroup is
generated by $\{B, K_0, V_-, W_+\}$.  From the commutation relations (2.2)
it appears that $G_0\equiv\,$U$(1/1)$, and thus the atypical $\OSp$-CS
are parametrized by the super K\"ahler homogeneous space
$\OSp/$U$(1/1)\equiv(\duu,\omega)$.  This could have been already deduced
from (2.5)--(2.6) and (\ref{W}).

Rothstein's data for $(\duu,\omega)$ are given in \cite{Amine1}.  They can
be rederived by simply using results of Section~4.  Indeed, the rank 2
holomorphic vector bundles $\E$ and $\F$ reduce both to a holomorphic line
bundle.  Moreover, from (\ref{metric}) one easily sees that when $b=-\tau$,
we are left with the Hermitian metric $||g||=g_{1\bar1}=2\tau/(\de)$.
This is twice the metric obtained in \cite{Amine1}.

On the other hand geometric quantization leads to the super Hilbert space
obtained in \cite{Amine2}.  This result corresponds to taking $b=-\tau$ in
Section~5.  Generators of $\osp$ are however represented by the following
operators,
\begin{eqnarray}
\widehat B{\phantom 1}&\!\!\!=\!\!\!&\displaystyle {\t\over 2 }\,
\pt   -\tau\ ,\qquad\qquad\qquad\quad\quad
\widehat K_0 = \displaystyle  z\, \pv+ {\t\over 2} \, \pt   +
\tau,\\
\widehat K_+ &\!\!\!=\!\!\!& \displaystyle  z^2 \, \pv+z\t \, \pt +2\tau z\ ,
\qquad\quad  \widehat K_- = \displaystyle \pv,\\
\widehat V_+ &\!\!\!=\!\!\!& \displaystyle -i z\t\, \pv -2i\tau\t   \ ,
\qquad\qquad\quad   \widehat W_- =  \displaystyle i\pt   \ , \\
\widehat V_- &\!\!\!=\!\!\!&
\displaystyle -i\t\, \pv  \ , \qquad\qquad\qquad\quad\quad\ \
\widehat W_+=
\displaystyle  iz \,  \pt \  .
\end{eqnarray}

Finally, the supermeasure on ${\cal D}^{(1\vert 1)}$ can not be obtained
simply as the limit $b\rightarrow-\tau$ of (\ref{medida}).  It has
to be evaluated starting from (\ref{s-form2}) using the same technique as
in Section~3.3.  This leads to,
\beq d\mu (z,\vb, \t,\tb)= {i\over \pi(\de)}\left[1+i\,{\tb\t \over
\de}\right] dz\, d\vb\, d\t\, d \tb\ .
\eeq

%%%%%%%%%%%%%%%%%%%%%%%%%%%%%%%%
%%%% 7. Miscellaneous results and discussions %%%%
%%%%%%%%%%%%%%%%%%%%%%%%%%%%%%%%
\section{Miscellaneous results and discussions}\label{sec7}

Here are gathered few consequences of the main results
of the paper.  Few other important points are discussed further.

\smallskip
\noindent{\bf 7.1.}
{\it Square integrability.\/} When speaking about coherent states the first
of their properties that comes to mind is the so-called resolution of the
identity.  In the non-super case this property reflects the square
integrability of the unitary irreducible representation these special states
belong to.  Does this notion   extend to the super case? The answer is yes.
A simple computation based on (\ref{medida}), (\ref{csb}) and the
Berezin integration leads to the following:
\beq
\int_{\dud} \ketb\brab|\,d\mu=\I. \label{resid}
\eeq
Here $\I\equiv\I_{\vtaub}$.  A similar identity holds for the atypical CS.
Hence, this allows a straightforward super extension of the definition of
a square integrable representation.  The above identity (\ref{resid})
provides a new argument that can be added to those already listed in
\cite{Nish} in order to justify  calling the super unitary irreducible
representations of $\OSp$ considered here {\it discrete series
representations\/}.  As for usual CS, another immediate consequence of
(\ref{resid}) is that
$\h_{\tau,b}$ is a reproducing super Hilbert space.  This applies to both
typical and atypical super Hilbert spaces.

\smallskip
\noindent{\bf 7.2.} {\it Status of $z$.\/} For simplicity the
variable $z$ was considered from the begining as a usual complex number
(see (\ref{z})).  The main reason behind this choice is to make the
connection between the  results of Sections~2 and 3, and Rothstein's
approach to supersymplectic supergeometry, as  described in Section~4,
free of any change of coordinates.  The same argument applies to the
integrals over the unit disc $\du$ that appear in Section~5 (see
(\ref{eq5.10})--(\ref{eq5.11})).  Indeed, if the soul of $z$ was different
from zero, then those integrals and the complex geometry of
Section~4 would be meaningless, unless a change of coordinates
transforming $z$ into a `soulless' variable is performed.  Hence, our initial
choice prevents us from making any change of coordinates.

\smallskip
\noindent{\bf 7.3.} {\it Whitney sum, Einstein-K\"ahler manifolds and
Einstein-Hermitian vector bundles.\/} The rank
$2$ holomorphic vector bundle $\E$ intervening in Rothstein's data for the
typical orbits (see Theorem~4.3)  is the Whitney sum of two
holomorphic line bundles over the unit disc
$\du$, i.e. $\E=\E_1\oplus \E_2$.  Independently, each component of $\E$
provides Rothstein's data for a super K\"ahler subsupermanifold of
$(\dud,\omega)$.  More precisely, $(\du,\omega_0,\E_i, g_i, \nabla^{g_i})$, for
$i=1$ or $2$ are such data for a $(1|1)$-dimensional super K\"ahler
supermanifold, denoted $\duu_{\tau\pm b}$, where $g_1$ (resp. $g_2$) is the
Hermitian metric on
$\E_1$ (resp. $\E_2$) given in (\ref{metric}) as the first (resp. second)
diagonal entry; and $\nabla^{g_i}$  are the associated Hermitian
connections.  These supermanifolds are $N=1$ extensions of
$(\du,\omega_0)$, the $2$-superforms of which are obtained from
(\ref{sform1})--(\ref{sform2}) by setting $\chi=0$ in the first instance
and $\t=0$ in the second.  These two super unit discs are  not
$\OSp$-homogeneous spaces, unless $b=\pm\tau$.  Indeed, when
$b=\pm\tau$ they both  become
$\duu$, the $N=1$ super unit disc, which is a supersymplectic
homogeneous space for both $\OSp$ and OSp$(1/2)$ as shown in Section~6.
Since both $\E_1$  and
$\E_2$ are trivial line bundles over $\du$, Rothstein's data for
$\duu_{\tau+ b}$ and $\duu_{\tau- b}$ differ only by the constant factor
${\tau\pm b}$ in front of the Hermitian structure on these bundles (see
(\ref{metric})).

The above observations
lead to the following interesting picture.  Let us denote by ${\bf
R}_\lambda$ the data $(\du,\omega_0,\E_\lambda, g_\lambda,
\nabla^{g_\lambda})$, where
$\omega_0$ is given in (\ref{O0}), $\E_\lambda$ a trivial holomorphic
line bundle over
$\du$, $||g_\lambda||=\lambda(\de)^{-1}$ (with $\lambda\geq0$) a
Hermitian structure on $\E_\lambda$, and $\nabla^{g_\lambda}$ the
corresponding Hermitian connection.  Then we have: {\bf (i)} when
$\lambda$ is a multiple of $\tau$, ${\bf R}_\lambda$ are just Rothstein's
data for $\duu$; {\bf (ii)} the other
$\OSp$-coadjoint orbit, $(\dud,\omega)$, admits as Rothstein's data ${\bf
R}_{\lambda_1}\oplus{\bf R}_{\lambda_2}$ such that
$\lambda_1+\lambda_2=2\tau$, where the symbol `$\oplus$'  indicates
that we have to take the Whitney sum of $(\E_{\lambda_1},
g_{\lambda_1})$ and
$(\E_{\lambda_2}, g_{\lambda_2})$.  We recall at this point that
the information in ${\bf R}_{\tau-b}\oplus{\bf R}_{\tau+b}$  are sufficient
to reconstruct the $2$-superform (\ref{sform1})--(\ref{sform2})  (see
Section~4).  Hence, ${\bf R}_{\lambda}$ is the basic building block for
describing super K\"ahler coadjoint orbits of simple Lie supergroups super
extending $\su$.

A deeper analysis of ${\bf R}_{\lambda}=(\du,\omega_0,\E_\lambda,
g_\lambda, \nabla^{g_\lambda})$ shows that it
is made of two main parts: {\bf (i)} $(\du,\omega_0)$, which is a
K\"ahler $\su$-homogeneous space, and {\bf (ii)} $(\E_\lambda,
g_\lambda)$, which is an {\it Einstein-K\"ahler manifold\/}
\cite{Koba}. Here we identify $\E_\lambda$ with the holomorphic
tangent bundle over $\du$.  On the other hand, it appears that
${\bf R}_{\lambda_1}\oplus{\bf R}_{\lambda_2}$ defines an
{\it Einstein-Hermitian vector bundle\/} \cite{Koba}.  These very
interesting and important observations not only improve the
characterization of a super K\"ahler coadjoint orbit, but they suggest a
way of extending to the super context the known classification of
irreducible bounded symmetric Hermitian domains.  This direction is now
under investigation.

\smallskip
\noindent{\bf 7.4.} {\it Realizations of the typical and atypical
representations.}  In agreement with the descriptions of Section~2
and Appendix~B, results of Section~5 show that the typical super Hilbert
space $\h_{\tau,b}$ of $L^2$ superholomorphic sections of $\lud$ is the
direct sum of Hilbert spaces of four holomorphic discrete series
representations of
$\su$.  More precisely, as a vector superspace
\beq
\h_{\tau,b}=\h_{k=\tau}\oplus 2\cdot
\h_{k=\tau+\demi}\oplus\h_{k=\tau+1},
\label{decomp}
\eeq
where $\h_k$ is the Hilbert space carrying the holomorphic discrete
series representation $D(k)$ of $\su$ (see (\ref{eq5.11}) and
(\ref{vtaub1})--(\ref{vtaub3})).  This suggests that the $\osp$ operators
obtained in (\ref{widehat}) and which act in the left hand side of
(\ref{decomp}) can be replaced by matrix valued and thus
anticommutating-variables free operators acting in the right hand side of
(\ref{decomp}) \cite{Amine3}.  The former realization is much more
convenient than the latter. Indeed, for example for osp$(N/2)$, the
matrices can be $2^N\times2^N$.  We insist here on the
fact that our main goal in Section~5 was to show that
geometric quantization
extends to the super context, at least when applied to
coadjoint orbits
admitting a super K\"ahler polarization.  We not only
succeeded in achieving this, but the above observation confirms that our
output constitutes an intrinsically supersymmetric alternative to
the matrix realization.  The same discussion applies to the atypical
representations.

\smallskip
\noindent{\bf 7.5.} {\it $\OSp$ representations.}
Throughout, we have been considering only representations
of the Lie
superalgebra.  Explicit representations of $\OSp$ can be in fact
obtained from
those of $\osp$ exhibited in Sections~5 and 6.  The procedure
does not present
any difficulties.  The first step towards this construction consists
in finding the
explicit action of $\OSp$ on
$\dud$ and $\duu$, by integrating the Hamiltonian vector fields in
(\ref{vectors}) and (6.10)--(6.13).  This amounts to solving super Riccati
differential equations, which have already been considered in \cite{Vero}.
The full construction will be given elsewhere.

%%%%%%%%%%%%%%%%%%%%%%%%%%%%%%%%%%%%%%
%%%% 8. Conclusions and future directions %%%%%%%
%%%%%%%%%%%%%%%%%%%%%%%%%%%%%%%%%%%%%%
\section{Conclusions and outlook}\label{sec8}

Although our present contribution treats a specific example,
the obtained
results pave the way to harmonic superanalysis.  It must be
regarded as
the first important step of a program aimed at classifying
Lie supergroups'
coadjoint orbits and the associated irreducible representations.

In this work several closely related questions have been addressed,
and
several new notions have been introduced.  The consistency of our
conventions is manifest throughout the paper, from abstract to
explicit representation theory via super K\"ahler geometry.  The main
results are now summarized:

\smallskip
\noindent{\bf (a)}  Starting with a comprehensive description
of the abstract typical and atypical representations of $\osp$, the
associated $\OSp$ coherent states are constructed.

\smallskip
\noindent{\bf (b)} Super extending known methods, their underlying
geometries are exhibited, and shown to be those of $\OSp$ coadjoint
orbits.  The latter are $\OSp$-supersymplectic homogeneous spaces:
$\dud\equiv\OSp/$(U$(1)\times$U$(1)$) for the typical CS, and
$\duu\equiv\OSp/$U$(1/1)$ for the atypical CS.

\smallskip
\noindent{\bf (c)} The identification of Rothstein's data for
$\dud$ and $\duu$, draws us to generalizing Rothstein's
theorem to the
complex-analytic setting.  This leads to a natural definition
of a super K\"ahler
supermanifold, $\dud$ and $\duu$ being  non-trivial examples
of such a  notion.
We moreover show that in this context, Rothstein's theorem
can be refined.  More
precisely, the {\it complete\/} supersymplectic structure of a
super K\"ahler
coadjoint orbit can be encoded in an elementary building
block of the type
mentioned in point 7.3 of the previous section.

\noindent {\bf (d)} Finally, geometric quantization is
successfully extended
to the super K\"ahler context exemplified by the typical and atypical
coadjoint orbits of $\OSp$.  A super K\"ahler polarization is exhibited in
each case. This leads to an explicit super unitary irreducible typical
(atypical) representation of
$\osp$ in a super Hilbert space of square integrable superholomorphic
sections of a complex line bundle sheaf over $\dud$ ($\duu$).

Possible generalizations of our results are numerous and worth
considering.  At both the
representation theoretic and the geometric levels, the present work
relies essentially on known results from the non-super context.  For
instance, the representation theory of
$\osp$ is based on that of
$\ssu\subset\osp$, while the supergeometry of the
$N=2$ (resp. $N=1$)
super unit disc $\dud$ (resp.
$\duu$) is based on that of the unit disc $\du$.
Our results show precisely
how the super extension occurs (see Section~7, 7.3).  One can now
seriously consider other Lie supergroups, and look for
a classification of
their super K\"ahler homogeneous spaces along the known
classification of
the K\"ahler homogeneous spaces of their body Lie groups.
Geometric
quantization will then provide a classification of their associated
representations.

%%%%%%%%%%%%%%%%%%%%%%%%%%%%%%%
%%%% Acknowledgements %%%%%%%%%%%%%%%
%%%%%%%%%%%%%%%%%%%%%%%%%%%%%%%

\paragraph*{Acknowledgments}
A.M.E. is indebted to S.T. Ali, C.
Duval, J. Harnad, and G. Tuynman for valuable and stimulating discussions.
He thanks the Mathematics Department of Concordia University, the Centre
de Recherches Math\'ematiques of Universit\'e de Montr\'eal, and the U.F.R.
de Math\'ematiques of Universit\'e de Lille for their kind hospitality, and
much appreciated support. L.M.N. acknowledges a fellowship from
Ministerio de Educaci\'on y  Ciencia (Spain) and kind hospitality at the
Centre de Recherches Math\'ematiques of Universit\'e de Montr\'eal.

%%%%%%%%%%%%%%%%%%%%%%%%%%%%%%%%%%%
%%%% Appendix A %%%%%%%%%%%%%%%%%%%%%%%%%
%%%%%%%%%%%%%%%%%%%%%%%%%%%%%%%%%%%%

\appendix
\section{Conventions and notation}\label{Appendix A}

One of the main features of the present work is the complete
consistency of
the conventions used throughout.  The latter are displayed here. They
concern the superalgebra ${\cal B}$,  introduced in Section~2, and its
interactions with both the Lie superalgebra and the Lie superalgebra's
modules.

A complex superalgebra is a complex vector superspace (i.e. a
$\Z_2$-graded linear space)
${\cal B}={\cal B}_0\oplus{\cal B}_1$ equipped with a
$\Z_2$-compatible product, namely,
${\cal B}_k\cdot{\cal B}_l\subset{\cal
B}_{k+l}$; ${\cal B}$ is considered  associative and possesses
a unity.  Note
that, ${\cal B}_0$ (resp.
${\cal B}_1$) is called the {\it even\/} (resp. {\it odd\/}) part of
${\cal B}$.  Accordingly, elements of
${\cal B}_0$ (resp. ${\cal B}_1$) are
called  {\it even\/} (resp. {\it odd\/}) elements of ${\cal B}$.  A
homogeneous element of ${\cal B}$ is either even or odd.
The {\it parity\/}
of such an element $u\in{\cal B}_k$, denoted $\epsilon(u)$, is defined by
$\epsilon(u)=k$.  The superalgebra ${\cal B}$ is supercommutative if
\beq  uv=(-1)^{\epsilon(u)\epsilon(v)}vu,
\eeq  for $u$ and $v$ two homogeneous elements of ${\cal B}$.

The {\it complex supercommutative superalgebra with unit\/} ${\cal B}$
considered in the present work is the complex Grassmann algebra
\cite{Berezin3,Cornwell} generated by $(\t,\chi)$ and their complex
conjugates $(\tb, \chib)$.  These are anticommuting, and hence nilpotent
variables. In other words ${\cal B}$  is the complex exterior algebra over
$\C^4=\C^2\oplus{\overline{\C}}^2$.  Its even (resp. odd) part is spanned by
the products of an even (resp. odd) number of generators, and the dimension
of
${\cal B}$ is $16$.  The decomposition of any element $\Theta\in{\cal B}$ in
a given basis of ${\cal B}$, assumes the following form
\beq
\Theta=\widetilde\Theta\cdot\hbox{I} + \Theta_{nil},
\eeq  where, the purely nilpotent component $\Theta_{nil}$ is called the {\it
soul\/} of $\Theta$, while the component $\widetilde\Theta$ along the
identity of ${\cal B}$ is called the {\it body\/} of $\Theta$.

The complex conjugation  maps $\C^2\ni(\t,\chi)\mapsto
(\tb,\chib)\in{\overline{\C}}^2$.  Its extension to ${\cal B}$ is completely
defined by the following rule:
\beq
\overline{\Theta_1\,\Theta_2}=\bar\Theta_1\,\bar\Theta_2,\qquad
\forall\ \Theta_1, \Theta_2\in {\cal B}.\label{conj}
\eeq
The other properties are:
\beq
\bar{\bar\Theta}=
\Theta\quad\hbox{and}
\quad\overline{\hbox{w}\,\Theta}={\bar{\hbox{w}}}\,\bar\Theta,
\quad \forall\ \hbox{w}\in \C\ \hbox{and}\ \Theta\in{\cal B}.
\eeq
An element $\Theta\in{\cal B}$ is {\it real\/} if $\bar\Theta=\Theta$.
Using  (\ref{conj}) one easily sees that $\Theta\bar\Theta$ is real for
$\Theta\in{\cal B}_0$ and imaginary for $\Theta\in{\cal B}_1$.

It is important to notice that our convention in (\ref{conj}) is different
from the one introduced by Berezin
\cite{Berezin2,Berezin3}, and commonly used in the literature (see
\cite{Schmitt} and references therein). In that case the complex
conjugate
of a product is taken as:
\beq
\overline{\Theta_1\,\Theta_2}=\bar\Theta_2\,\bar\Theta_1\label,\qquad
\forall\ \Theta_1, \Theta_2\in {\cal B}.\label{Bconj}
\eeq Hence, for $\Theta$ a homogeneous element of ${\cal B}$,
$\Theta\bar\Theta$ is real independently of the parity of $\Theta$.
Using
these conventions one faces serious inconsistencies.  The most
obvious one
was encountered in
\cite{Amine1} (see also \cite{BKLR}), where the author followed
Berezin's conventions already used in \cite{Balal1}; the super K\"ahler
$2$-form obtained there was neither real nor imaginary!   As a
consequence, the classical observables and their associated Hamiltonian
vector fields were not satisfying any property of the type of
(\ref{reality}) and (\ref{reality'}), which are crucial in identifying a real
observable and then the associated self superadjoint operator.  When
combined with the notion of a super Hermitian structure of
\cite{SternWolf}, our convention (\ref{conj}) cures this discrepancy.
More precisely, the arguments invoked in \cite{Schmitt} in order to
justify the choice in (\ref{Bconj}) apply to our choice (\ref{conj}) too,
provided  one considers the notion of a super Hermitian structure
(\ref{SH1})--(\ref{SH3}) and its consequences.
(The problem in
\cite{Amine1} mentioned above is cured in \cite{Amine2}).

Finally, all vector superspaces appearing in this work are
considered as left
${\cal B}$-modules.  Let $V=V_{\bar 0}\oplus V_{\bar1}$
be such a
${\cal B}$-module.  Then, for $v$ and $\Theta$ homogeneous
elements in
$V$ and ${\cal B}$ respectively, we have,
\beq
\Theta\, v=(-1)^{\epsilon(\Theta)\epsilon(v)}v\,\Theta.
\eeq
This applies equally well to $V=\ospc$, $V=\vtaub$ or
$V=U(\pm\tau)$.  When $V$ is equipped with an additional
structure, such
as a super Hermitian form $\langle\, \cdot\, |\, \cdot\,\rangle$ (see
(\ref{SH1})--(\ref{SH2'})) or a Lie superalgebra bracket $[\,,\,]$,
then that
structure can be extended to the {\it Grassmann envelope of
second type\/}
$\widetilde V$ of
$V$.  The latter is defined as follows
\cite{Berezin3},
\beq
\widetilde V\equiv\left({\cal B}\otimes V\right)_0=\left({\cal
B}_0\otimes V_0\right)\oplus\left({\cal B}_1\otimes V_1\right),
\eeq and the structures above are extended in the following way:
\beq
\langle\Theta_1\,v|\Theta_2\,
u\rangle=(-1)^{\epsilon(v)\epsilon(\Theta_2)}\bar\Theta_1\Theta_2
\langle v| u\rangle \quad \hbox{and}\quad \left[\Theta_1\,X,
\Theta_2\,
Y\right]=(-1)^{\epsilon(X)\epsilon(\Theta_2)}\Theta_1\Theta_2\left[X,Y\right],
\eeq where $\Theta_2$, $v$ and $X$ are homogeneous
elements in respectively, ${\cal B}$, $\vtaub$ and $\ospc$.
We end this
appendix by giving a formula which is useful for some of the
computations
of Section~2 and Appendix~B.  Let $\Theta$ and $v$ be
homogeneous elements
of respectively $\vtaub$ and ${\cal B}$, the super
Hermitian conjugate of
$|\Theta\,v\rangle\in\widetilde{V}(\tau,b)$ with
respect to the super
Hermitian form (\ref{SH2})--(\ref{SH3}), is
obtained as follows:
\beq
\left(\Theta\,|v\rangle\right)^{\dagger}=
\bar\Theta\,\left(|v\rangle\right)^{\dagger}
= (i)^{\epsilon(v)}\bar\Theta\, \langle v|\label{sHconj}.
\eeq

%%%%%%%%%%%%%%%%%%%%%%%%%%%%%%%%%%%%
%%%% Appendix B %%%%%%%%%%%%%%%%%%%%%%%%%
%%%%%%%%%%%%%%%%%%%%%%%%%%%%%%%%%%%%

\section{osp(2/2) representations: more details}\label{Appendix B}

Finite and infinite dimensional irreducible representations of \osp\ have
already  been studied in \cite{ShNaRitt,Nish,Balal2}.
The description of the infinite dimensional ones given in \cite{Balal2} is
the most convenient for our purpose, but since it suffers from some
discrepancies we consider important to reexpose the construction.  This
appendix must be viewed as a complement to Section~2.

The equations defining the lowest weight vector $\ko$
(\ref{2.7})--(\ref{2.8}), together with the observation that $\ko$ is
the lowest weight vector of a discrete series representation $D(\tau)$ of
$\ssu$ (\ref{discrete}), are our starting points.

The $\osp$-module $\vtaub$ of Section~2 is generated by applying
arbitrary polynomials in the generators of
$\n^+=\hbox{span}\{K_+,V_+,W_+\}$ (\ref{n}) to
$\ko$.  Using
the commutation relations and the results mentioned above,
one can see that
$\vtaub$ is spanned by the following vectors:
\beq
{K}_+^m\ko,\quad {K}_+^m{  V}_+\ko,\quad   {K}_+^m{W}_+\ko, \quad
{K}_+^m{  V}_+ {W}_+\ko,\quad m\in\N. \label{B.4}
\eeq
The latter are eigenstates of $B$:
\begin{eqnarray}
B(K_+^m \ko)&\!\!\!=\!\!\!&b(K_+^m\ko) \nonumber \\
B(K_+^m V_+\ko)&\!\!\!=\!\!\!&(b+\textstyle{1\over2}) (K_+^m V_+\ko)
\nonumber\\
B(K_+^m W_+\ko)&\!\!\!=\!\!\!&(b-\textstyle{1\over2}) (K_+^m W_+\ko)
\nonumber\\
B(K_+^mV_+W_+\ko)&\!\!\!=\!\!\!&b (K_+^m V_+W_+\ko),
\end{eqnarray}
and also of $K_0$:
\begin{eqnarray}
K_0(K_+^m\ko)&\!\!\!=\!\!\!&(\tau+m)(K_+^m\ko) \nonumber\\
K_0(K_+^m V_+\ko)&\!\!\!=\!\!\!&(\tau+\textstyle{1\over2}+m) (K_+^m
V_+\ko) \nonumber\\
K_0(K_+^m W_+\ko)&\!\!\!=\!\!\!&(\tau+\textstyle{1\over2}+m) (K_+^m
W_+\ko) \nonumber\\
K_0(K_+^m V_+W_+\ko)&\!\!\!=\!\!\!&(\tau+1+m) (K_+^mV_+W_+\ko),
\end{eqnarray}
but not all of them are eigenstates of  the $\ssu$ Casimir
$C_2$ given in (\ref{C2}):
\begin{eqnarray}
C_2(K_+^m\ko)&\!\!\!=\!\!\!&\tau(\tau-1)(K_+^m\ko) \nonumber\\
C_2(K_+^m
V_+\ko)&\!\!\!=\!\!\!&(\tau+\textstyle{1\over2}) (\tau-\textstyle{1\over2})
(K_+^m V_+\ko) \nonumber\\
C_2(K_+^m W_+\ko)&\!\!\!=\!\!\!&(\tau+\textstyle{1\over2})
(\tau-\textstyle{1\over2}) (K_+^m W_+\ko) \nonumber\\
C_2(K_+^m
V_+W_+\ko)&\!\!\!=\!\!\!&(\tau+1)\tau
(K_+^m V_+W_+\ko)-(\tau+b)(K_+^{m+1}\ko).
\end{eqnarray}
The last equation suggests to use another family of
states instead of $K_+^m
V_+ W_+\ko$. Indeed, notice that the vectors
\beq
\left(2\tau K_+^m V_+ W_+ -(\tau+b) K_+^{m+1}\right)\ko,\ \
\left(\hbox{or}\
\
\left(-2\tau K_+^m W_+ V_+ +(\tau-b) K_+^{m+1}\right)\ko\right),
\eeq
are such that
\begin{eqnarray}
B\left(2\tau{  K}_+^m{  V}_+{  W}_+ -(\tau+b)
K_+^{m+1}\right)\!\!\ko&\!\!\!\!=\!\!\!\!&b \left(2\tau{  K}_+^m V_+{  W}_+
-(\tau+b){  K}_+^{m+1}\right)\!\!\ko  \\
C_2\left(2\tau{  K}_+^m{  V}_+{  W}_+
-(\tau+b){  K}_+^{m+1}\right)\!\!\ko&\!\!\!\!=\!\!\!\!&(\tau+1)\tau
\left(2\tau{
K}_+^m{  V}_+{  W}_+-(\tau+b){  K}_+^{m+1}\right)
\!\!\ko \nonumber \\
{  K}_0\left(2\tau{  K}_+^m {  V}_+{  W}_+ -(\tau+b){
K}_+^{m+1}\right)\!\!\ko&\!\!\!\!=\!\!\!\!&(\tau+1+m)\!
\left(2\tau{  K}_+^m{  V}_+{  W}_+ -(\tau+b){  K}_+^{m+1}\right)\!\!\ko.
\nonumber
\end{eqnarray}
The previous results suggest to use the following  notation:
\begin{eqnarray}
 |b,\tau,\tau\rangle\equiv\ko, \qquad\qquad\qquad\qquad\qquad\, &
|b+\textstyle{1\over2},\tau+\textstyle{1\over2},
\tau+\textstyle{1\over2}\rangle\propto {  V}_+\ko,\ \, \nonumber\\
|b,\tau+1,\tau+1\rangle\propto
\left(2\tau{  V}_+{ W}_+ -(\tau+b){  K}_+\right)\!\ko,
&  |b-\textstyle{1\over2},\tau+\textstyle{1\over2},
\tau+\textstyle{1\over2}\rangle\propto {  W}_+\ko. \label{B.10}
\end{eqnarray}

The $\osp$-module $\vtaub$ obtained in this way is irreducible only when
$b\not=\pm\tau$.  For $b=\pm\tau$, $\vtaub$ contains a {\it primitive
vector\/}.   Indeed, using (2.2a)--(2.2i) and (\ref{2.7})--(\ref{2.8}), one
easily sees that:
\beq
\hbox{when}\quad b=\tau, \qquad
K_-(V_+\ko)=V_-(V_+\ko)=W_-(V_+\ko)=0,
\eeq
and
\beq
\hbox{when}\quad b=-\tau, \qquad
K_-(W_+\ko)=V_-(W_+\ko)=W_-(W_+\ko)=0.
\eeq
These two situations being very similar, we focus here only on the second
one.   Hence, in that instance, $W_+\ko$ generates an
$\osp$-submodule of $\vtaub$, denoted $V'(\tau,-\tau)$.  An irreducible
$\osp$-module emerges then as the quotient
$V(\tau,-\tau)/V'(\tau,-\tau)\equiv U(-\tau)$.  Notice
that $V'(\tau,-\tau)$ is spanned by the last two series in (\ref{B.4}),
while $U(-\tau)$ is spanned by the two first ones modulo the two last
ones.

In order to obtain the proportionality constants in (\ref{B.10}) it is
necessary to equip $\vtaub$ with a super Hermitian form
\cite{SternWolf,Scheunert} of the type described in
(\ref{SH1})--(\ref{SH2'}).  The {\it superadjoint\/} $A^{\dagger}$
(denoted $\tilde A$ and called differently in \cite{Scheunert}) of a
homogeneous operator
$A$ acting in $\vtaub$ is defined as follows:
\beq
\langle A^{\dagger}u|v\rangle=(-1)^{\epsilon(u)\epsilon(A)}\langle
u|Av\rangle, \quad \forall\ u,
v\in\vtaub\quad\hbox{with}\ u\ \hbox{homogeneous}.\label{superA}
\eeq
Hence, $A$ is self superadjoint if $A^{\dagger}=A$.  Moreover, one can
check from  (\ref{superA}) that
\beq
(A^{\dagger})^{\dagger}=A\qquad\hbox{and}\qquad
(AB)^{\dagger}=(-1)^{\epsilon(A)\epsilon(B)} B^{\dagger} A^{\dagger}.
\eeq
Consequently,
\beq
\left[A,B\right]^{\dagger}=-\left[A^{\dagger},B^{\dagger}\right].
\eeq
Using this equation, the relations in (2.2a)--(2.2i) and
the root space
decomposition of $\ospc$ given in Section~2, one
easily shows that
\beq
B^{\dagger}= B,\ \  \ K_0^{\dagger}=K_0,
\ \  \ ({K}_{\pm})^{\dagger}={K}_{\mp},
\label{superA1}
\eeq
\beq
(V_{\pm})^{\dagger}=i W_{\mp}\quad\hbox{and}\quad
(W_{\pm})^{\dagger}=i V_{\mp}\ . \label{superA2}
\eeq

Using the above results, equation (\ref{sHconj}), and assuming that the
vectors on the left hand side of the equations in (\ref{B.10}) are
normalized
to one with respect to
$\langle\,\cdot\,|\,\cdot\,\rangle_{0(1)}$ (according to their parity) one
obtains:
\beq
{  V}_+ |b,\tau,\tau\rangle=\sqrt{\tau-b}\,|b+{\textstyle{1\over2}},\tau+
{\textstyle{1\over2}},\tau+{\textstyle{1\over2}}\rangle
\ ,\label{B15}
\eeq
\beq
{  W}_+ |b,\tau,\tau\rangle=\sqrt{\tau+b}\,|b-{\textstyle{1\over2}},
\tau+{\textstyle{1\over2}},\tau+ {\textstyle{1\over2}}\rangle
\  ,
\eeq
\beq
\left(2\tau{  V}_+{  W}_+ -(\tau+b){  K}_+\right)  |b,\tau,\tau\rangle=
\sqrt{(\tau^2-b^2) (2\tau+1)2\tau}\,|b,\tau+1,\tau+1\rangle\ .  \label{B17}
\eeq
Clearly, these results are valid only if $|b|\leq\tau$.  When
$b\not=\pm\tau$, this means that the super Hermitian structure
introduced above, turns the irreducible $\osp$-modules $\vtaub$ into
super unitary representations only if $|b|<\tau$. On the other hand, when
$b=\pm\tau$ one sees from (\ref{B15})--(\ref{B17}) that the primitive
vectors generating the submodules $V'(\tau,\pm\tau)$ are zero-norm
states, and thus the entire submodules are made of zero-norm states.
Moding out the latter from
$V(\tau,\pm\tau)$ turns $U(\pm\tau)$ into super unitary irreducible
modules.    Moreover, (\ref{vtaub2}) and (\ref{vtaub4}) are direct
consequences of (\ref{B15})--(\ref{B17}) and (2.2a)--(2.2i).  Indeed,
the four states
$|b,\tau,\tau\rangle$, $V_+|b,\tau,\tau\rangle$,
$W_+|b,\tau,\tau\rangle$ and $\left(2\tau{  V}_+{ W}_+ -(\tau+b){
K}_+\right)\!|b,\tau,\tau\rangle$ are orthogonal with respect to the super
Hermitian structure, and each of them is a lowest state of an $\ssu$
irreducible module.  The latter are generated from the former
$\ssu$ lowest-weight states through the action of powers of $K_+$:
\beq
{  K}_+^m|\,\cdot\,,k,k\rangle=\sqrt{ {m!\,
\Gamma(2k+m)\over
\Gamma(2k)}}\,|\,\cdot\,,k,k+m\rangle;\quad
k=\tau,\tau\pm\textstyle{1\over2}, \hbox{or}\ \tau+1.  \label{B18}
\eeq

We end this appendix by displaying the action of the $\osp$
generators on
the different vectors.  The following formulae are straightforward
consequences of (2.2a)--(2.2i) and the results described above;
\begin{eqnarray}
{K}_+|b,\tau,\tau+m\rangle &=&\sqrt{(2\tau+m)(m+1)}\,
|b,\tau,\tau+m+1\rangle ,  \nonumber\\
{K}_-|b,\tau,\tau+m\rangle &=&\sqrt{(2\tau+m-1)m}\,
|b,\tau,\tau+m-1\rangle ,   \nonumber\\
{V}_+|b,\tau,\tau+m\rangle &=&\sqrt{\textstyle{(\tau-b)(2\tau+m)\over
2\tau}}
\,|b+\textstyle{1\over2},\tau+\textstyle{1\over2},\tau+
\textstyle{1\over2}+m\rangle ,   \nonumber \\
{V}_-|b,\tau,\tau+m\rangle &=&\sqrt{\textstyle{(\tau-b)m\over 2\tau}}
\,|b+\textstyle{1\over2},\tau+\textstyle{1\over2},\tau+\textstyle{1\over2}+
m-1\rangle ,   \nonumber \\
{W}_+|b,\tau,\tau+m\rangle &=&\sqrt{\textstyle{(\tau+b)(2\tau+m)\over
2\tau}}
\,|b-\textstyle{1\over2},\tau+\textstyle{1\over2},\tau+
\textstyle{1\over2}+m\rangle ,   \nonumber \\
{W}_-|b,\tau,\tau+m\rangle &=&
\sqrt{\textstyle{(\tau+b)m\over 2\tau}}
\,|b-\textstyle{1\over2},\tau+\textstyle{1\over2},
\tau+\textstyle{1\over2}+
m-1\rangle ,  \nonumber\\
{V}_+|b+\textstyle{1\over2},\tau+\textstyle{1\over2},\tau+
\textstyle{1\over2}+m\rangle &=& 0 ,  \nonumber\\
{V}_-|b+\textstyle{1\over2},\tau+\textstyle{1\over2},\tau+
\textstyle{1\over2}+m\rangle &=& 0 ,  \nonumber\\
{W}_+|b+\textstyle{1\over2},\tau+\textstyle{1\over2},\tau+
\textstyle{1\over2}+m\rangle &=&
\sqrt{\textstyle{(\tau-b)(m+1)\over 2\tau}}
\,|b,\tau,\tau+m+1\rangle   \nonumber\\
\phantom{{ W}_+|b+\textstyle{1\over2},\tau+\textstyle{1\over2},
                \tau+\textstyle{1\over2}+m\rangle=}
& & - \sqrt{\textstyle{(\tau+b)(2\tau+m+1)\over 2\tau}}
\,|b,\tau+1,\tau+1+m\rangle ,   \nonumber\\
{W}_-|b+\textstyle{1\over2},\tau+\textstyle{1\over2},\tau+
\textstyle{1\over2}+m\rangle &=&
\sqrt{\textstyle{(\tau-b)(2\tau+m)\over 2\tau}}
\,|b,\tau,\tau+m\rangle   \label{Action}\\
   \phantom{{ W}_+|b+\textstyle{1\over2},\tau+\textstyle{1\over2},
                \tau+\textstyle{1\over2}+m\rangle=}
& &-\sqrt{\textstyle{(\tau+b)m\over 2\tau}}
\,|b,\tau+1,\tau+1+m-1\rangle ,   \nonumber\\
{V}_+|b-\textstyle{1\over2},\tau+\textstyle{1\over2},\tau+
\textstyle{1\over2}+m\rangle
&=&\sqrt{\textstyle{(\tau+b)(m+1)\over 2\tau}}
\,|b,\tau,\tau+m+1\rangle    \nonumber\\
  \phantom{{ W}_+|b+\textstyle{1\over2},\tau+\textstyle{1\over2},
                \tau+\textstyle{1\over2}+m\rangle=}
& &+ \sqrt{\textstyle{(\tau-b)(2\tau+m+1)\over 2\tau}}
\,|b,\tau+1,\tau+1+m\rangle ,   \nonumber\\
{V}_-|b-\textstyle{1\over2},\tau+\textstyle{1\over2},\tau+
\textstyle{1\over2}+m\rangle &=&
\sqrt{\textstyle{(\tau+b)(2\tau+m)\over 2\tau}}
\,|b,\tau,\tau+m\rangle    \nonumber\\
    \phantom{{ W}_+|b+\textstyle{1\over2},\tau+\textstyle{1\over2},
                \tau+\textstyle{1\over2}+m\rangle=}
& &+ \sqrt{\textstyle{(\tau-b)m\over 2\tau}}
\,|b,\tau+1,\tau+1+m-1\rangle ,   \nonumber\\
{W}_+|b-\textstyle{1\over2},\tau+\textstyle{1\over2},\tau+
\textstyle{1\over2}+m\rangle &=&0 ,  \nonumber\\
{W}_-|b-\textstyle{1\over2},\tau+\textstyle{1\over2},\tau+
\textstyle{1\over2}+m\rangle &=&0 ,  \nonumber\\
{V}_+|b,\tau+1,\tau+1+m\rangle &=&-\sqrt{\textstyle{
(\tau+b)(m+1)\over 2\tau}}
\,|b+\textstyle{1\over2},\tau+\textstyle{1\over2},\tau+\textstyle
{1\over2}+m+ 1\rangle ,   \nonumber\\
{ V}_-|b,\tau+1,\tau+1+m\rangle &=&-\sqrt{
\textstyle{(\tau+b)(2\tau+m+1)\over 2\tau}}
\,|b+\textstyle{1\over2},\tau+\textstyle{1\over2},\tau+
\textstyle{1\over2}+m\rangle ,   \nonumber\\
{W}_+|b,\tau+1,\tau+1+m\rangle &=&
\sqrt{\textstyle{(\tau-b) (m+1)\over 2\tau}}
\,|b-\textstyle{1\over2},\tau+\textstyle{1\over2},\tau+
\textstyle{1\over2}+m+1\rangle ,   \nonumber\\
{W}_-|b,\tau+1,\tau+1+m\rangle &=&
\sqrt{\textstyle{(\tau-b) (2\tau+m+1)\over
2\tau}}
\,|b-\textstyle{1\over2},\tau+\textstyle{1\over2},
\tau+\textstyle{1\over2}+m\rangle .   \nonumber
\end{eqnarray}
The discrepancies mentioned at the begining of this
appendix can be easily seen by comparing
(\ref{Action}) with its analog in
\cite{Balal2}.

%%%%%%%%%%%%%%%%%%%%%%%%%%%%%%%%%%
%%%% References %%%%%%%%%%%%%%%%%%%%%%
%%%%%%%%%%%%%%%%%%%%%%%%%%%%%%%%%
%{\sect\noindent
%References}
%\vglue 0.4cm
{\footnotesize{

}}

\end{document}